\documentclass[aps,preprint]{revtex4-1}

\draft 
\usepackage{graphicx}
\usepackage{subcaption}

\usepackage{bm}
\usepackage{dcolumn}
\usepackage{ascmac}
\usepackage{amsmath,amssymb}

\newcommand{\mathbd}[1]{\mbox{\boldmath ${#1}$}}
\newcommand{\pd}[2]{{\frac{\partial {#1}}{\partial {#2}}}}

\newcommand{\pdpd}[3]{{\frac{\partial^2 {#1}}{\partial {#2}\partial {#3}}}}

\newcommand{\avrgxz}[1]{{\left\langle {#1} \right\rangle_{xz}}}

\begin{document}


\title{Searching for turbulence models by artificial neural network} 



\author{Masataka Gamahara}
\affiliation{Graduate School of Information Sciences, Tohoku University, Sendai 980-8579, Japan}
\author{Yuji Hattori}
\email[]{hattori@fmail.ifs.tohoku.ac.jp}
\thanks{corresponding author}
\affiliation{Institute of Fluid Sciences, Tohoku University, Sendai 980-8577, Japan}


\date{\today}

\begin{abstract}
Artificial neural network (ANN) is tested as a tool for finding a new subgrid model 
of the subgrid-scale (SGS) stress in large-eddy simulation. 
ANN is used to establish a functional relation between the grid-scale (GS) flow field 
and the SGS stress without any assumption of the form of function. 
Data required for training and test of ANN are provided by 
direct numerical simulation (DNS) of a turbulent channel flow. 
It is shown that ANN can establish a model similar to the gradient model. 
The correlation coefficients between the real SGS stress and the output of ANN 
are comparable to or larger than similarity models, 
but smaller than a two-parameter dynamic mixed model. 
\end{abstract}


\maketitle 

\section{Introduction}

Large-eddy simulation (LES) is used as an important tool 
of numerical simulation in a wide variety of fields where 
turbulent flows appear. 
In LES the turbulent flow fields are decomposed into 
resolved-scale or grid scale (GS) flow field 
and small-scale or subgrid-scale (SGS) fluctuations 
by a filtering operation. 
In incompressible turbulent flows 
the effects of the fluctuations on the GS flow field 
appear as the residual or SGS stress tensor.   
How to model the SGS stress tensor using the GS flow field 
is the most important issue in LES. 

A number of subgrid models have been proposed since the Smagorinsky model \cite{S-1963}.  
Most of them are categorized into 
the Smagorinsky model,  
the similarity model \cite{BFR-1984}, 
and the gradient model \cite{CFR-1979};  
there are also dynamic versions of these models \cite{G-1992, VGK-1996} and mixed models \cite{ZSK-1993};  
see e.g. Lesieur and M\'etais \cite{LM-1996} and Meneveau and Katz \cite{MK-2000} for reviews. 
The performance of these models have been tested for some particular flows like  
isotropic turbulence \cite{FTWG-1997} and mixing layer \cite{VGK-1997}.   
There is no model which is better than the other models in any flows, 
although the dynamic and mixed models often give more accurate results than non-dynamic models.  
Accuracy of the results obtained by LES is still limited 
as they should be interpreted with much care 
when experimental or DNS results are unavailable for validation;  
controversially, however, this is the very situation in which LES is most wanted. 
Thus, a new subgrid model which is much better than the existing ones should be developed. 
In order to pursue such a subgrid model  
essentially new ideas of modeling would be required;  
however, there has been no such idea 
since the prototypes of the above models were proposed. 

The final goal of our research is to establish a new subgrid model 
for the SGS stress 
which performs better than the existing models. 
This is not an easy task 
since there are potentially a huge number of possibilities of modeling; 
in principle one can use the GS flow field in the whole domain to express the SGS stress at one point 
as in nonlocal methods in RANS modeling \cite{H-2005}; 
history of the GS flow field may also be used as in the Reynolds stress transport model \cite{P-2002}. 
On the other hand, 
the subgrid model should be based on physics of turbulence 
and it seems difficult to invent an essentially new model 
after the models listed above were established with much efforts. 
However, the methods of machine learning, 
which are extensively and successfully used in many areas, 
can provide us a tool for going beyond the ability of human thoughts;  
they are useful if they can automatically extract 
the essential GS flow field required for accurate modeling. 
In this paper, 
we test whether the artificial neural network (ANN), 
which is one of the methods of machine learning, 
can be used for finding a subgrid model;  
it is a first step toward the above goal.  

There have been many applications of ANN to turbulence modeling. 
Sarghini {\textit{et al.}} \cite{SdFS-2003} used ANN to find a relation between the GS flow field 
and the turbulent viscosity coefficient in the mixed model. 
Moreau {\textit{et al.}} \cite{MTB-2006}
used ANN to estimate the subgrid variance in 
the Cook and Riley model for a scalar field in isotropic turbulence.  
Recently ANN was used in Ma {\textit{et al.}} \cite{MLT-2015}
to find closure terms for a one-dimensional model of bubbly multiphase flows. 
There are many applications of ANN to combustion \cite{ISP-2009, SM-2009, SHM-2010, EF-2012};  
ANN is used to speed up sub-grid chemistry computations, 
which are usually the bottle-neck of LES of turbulent flames.  
In plasma turbulence Citrin {\textit{et al.}} \cite{Cetal-2015}
used ANN to construct a model for tokamak turbulence transport model. 
In most of these works, however, ANN was used only as a complementary tool 
for optimizing model constants. 
Moreover, few works have dealt with SGS stress arising from convective terms,  
which is intrinsic to turbulence. 
In other words, there has been no attempt to construct an essentially new subgrid model 
of the SGS stress using ANN. 

It should be pointed out that 
we aim at establishing a functional relation between 
the GS flow field and the SGS stress tensor 
without any assumption on its form;  
this has not been done in the past. 
It is contrasted to the approach by Sarghini {\textit{et al.}} \cite{SdFS-2003}, 
who assumed the relation be the mixed model; 
their aim was to reduce the computational time by 
replacing the evaluation of the turbulent viscosity coefficient by ANN; 
the training target was the Bardina model instead of the real SGS stress. 
If ANN is shown to be an effective tool for turbulence modeling, 
it can be also applied to 
complex flows including compressible flows, multi-phase flows, and reactive flows,  
in which it is not always clear how to model the residual terms.  

The paper is organized as follows.  
Numerical methods are described in \S 2; 
after describing the outline in \S 2.1, 
methods of DNS are detailed in \S 2.2; 
preparation of data for ANN is summarized in \S 2.3; 
methods of ANN including the choice of input variables are described in \S 2.4. 
The results are shown in \S 3; 
first, choice of input variables is discussed in \S 3.1; 
in \S 3.2 ANN is shown to be successful in learning the SGS stress; 
basic features of learning are shown in \S 3.3; 
applicability of ANN trained at a low Reynolds number to higher Reynolds numbers 
is tested in \S 3.4; 
finally, what kind of model ANN has established is investigated in \S 3.5. 
Summary and future works are given in \S 4.

\section{Numerical Methods}

\subsection{Outline}

In LES small fluctuations of a flow variable $f$ are filtered out 
and we are concerned with the resolved-scale or GS flow field  
$\overline{f} = \int G(\pmb{x}') f (\pmb{x}-\pmb{x}') d\pmb{x}'$, 
where $G$ is a filter function.  
The GS flow field is governed by the filtered Navier-Stokes equations 
\begin{eqnarray}
\pd{\overline{u_i}}{t} + \pd{}{x_j} \left(\overline{u_i} \ \overline{u_j}\right) 
&=& -\pd{\overline{p}}{x_i} + \frac{1}{Re} \pdpd{\overline{u_i}}{x_k}{x_k} - \pd{\tau_{ij}}{x_j}, \\
\pd{\overline{u_j}}{x_j} &=& 0.  
\end{eqnarray}
Here the residual or SGS stress tensor  
\begin{eqnarray}
\tau_{ij} = \overline{u_i u_j} -\overline{u_i} \ \overline{u_j} 
\end{eqnarray}
depends not only on the GS flow field but also on fluctuations 
and should be modeled using the GS flow field. 

In the present study we use ANN to 
establish a functional relation between 
the GS flow field and the SGS stress tensor.  
First, training data for ANN are prepared using DNS data; 
the GS flow field are calculated from DNS data and used as input variables of ANN; 
the SGS stress tensor is also calculated from DNS data and used as training targets of output variables. 
Then, ANN is trained to establish a functional relation between the input and output variables. 
Finally, the ability of the trained ANN is checked 
using DNS data which are not used in training.  

\subsection{Direct numerical simulation}

The data used for training and test are obtained by DNS of
a turbulent channel flow. 
We use the formulation of Kim {\textit{et al.}} \cite{KMM1987}. 
For spatial discretization
the sixth-order accurate compact scheme is used in $y$ which is the direction
normal to the walls,
while the Fourier collocation method is used in the streamwise direction $x$ 
and the spanwise direction $z$ 
assuming periodic boundary conditions. 
Non-uniform grids are adopted in $y$ to resolve the boundary layers. 
The Poisson equations are solved in the Fourier space
where they are decomposed into
independent second-order ordinary differential equations for Fourier modes; 
they are also discretized by the compact scheme and solved accurately
and efficiently. 
The friction Reynolds number is $Re_\tau =u_\tau \delta/\nu= 180, 400, 600$ and $800$,
where $u_\tau=\tau_w^{1/2}$, $\tau_w=(1/Re) \pd{u}{y}$, and $\delta=1$ is the channel half-width.
The size of numerical domain $L_x\times L_y\times L_z$, 
the grid spacings in wall unit 
$\Delta x^+$, $\Delta y^{+}_{\rm{max}}$, and $\Delta z^{+}$, 
and the number of grid points $N_x\times N_y\times N_z$ 
are listed in Table \ref{table-grid}.

\begin{table}[h]
	\begin{center}
    \vspace{5mm}
		\caption{Parameter values of DNS. }
		\def~{\hphantom{0}}
		\begin{tabular}{cccccccc}
\hline\hline
			$Re_\tau$ & $L_x$ & $L_y$ & $L_z$ & $\Delta x^+$ & $\Delta y^{+}_{\rm{max}}$ & $\Delta z^{+}$ & $N_x \times N_y \times N_z $\\ \hline
			180&$4\pi$~&2~&2$\pi$~&11.8~&5.4~&7.1~&$192\times128\times160$\\
		 	400&$2\pi$~&2~& $\pi$~& 9.8~&7.9~&4.9~&$256\times192\times256$\\
			600&$2\pi$~&2~& $\pi$~& 7.4~&8.1~&3.7~&$512\times256\times512$\\
			800&$ \pi$~&2~& $\pi$~& 6.5~&7.7~&3.3~&$384\times384\times768$\\
\hline\hline
  		\end{tabular}
  		\label{table-grid}
  	\end{center}
\end{table}

\begin{figure}[h]
	\centerline{\includegraphics[width=70mm]{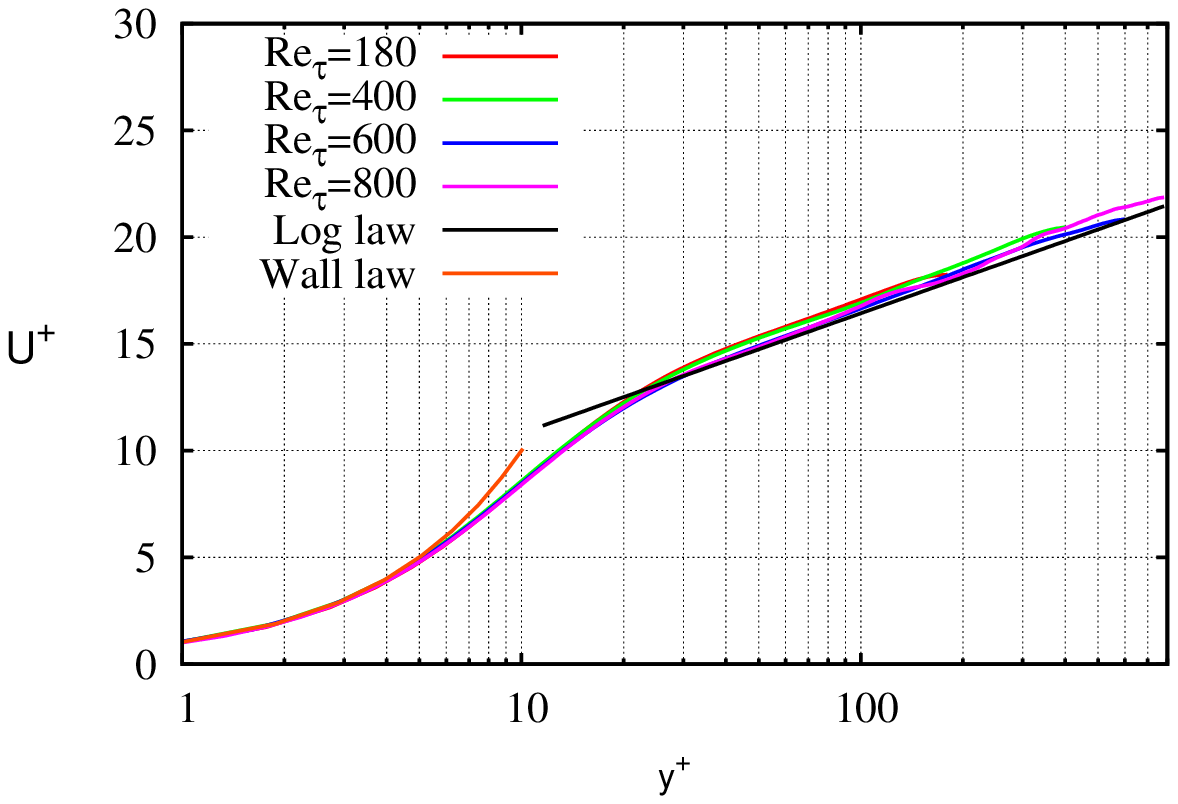}}   
	\centerline{\includegraphics[width=70mm]{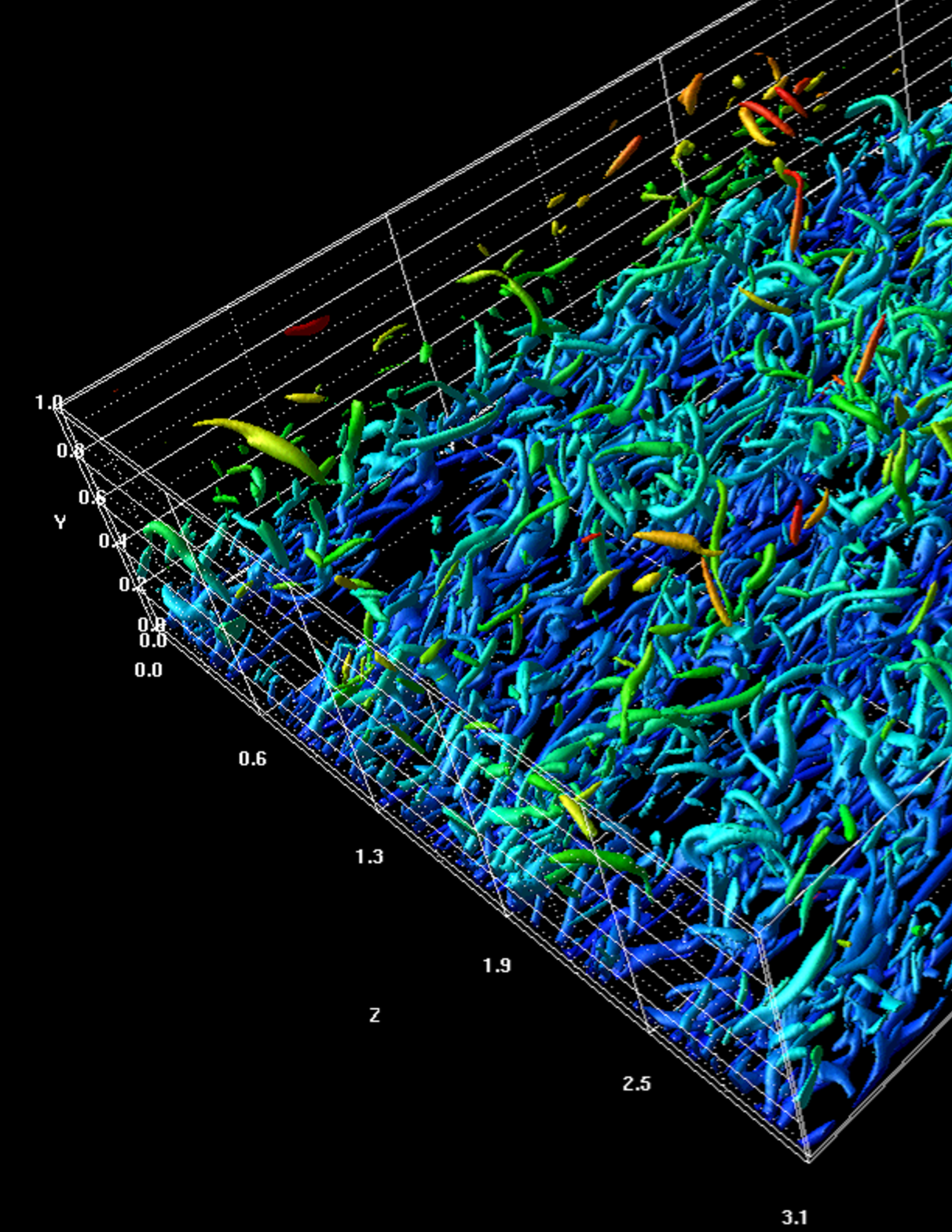}}
  \caption{Results of DNS of channel flow. 
(a) Mean flow at $Re_\tau =180, 400, 600$, and $800$;  
(b) vortical structures shown by iso-surface of the second invariant $Q$ of the deformation tensor. $Re_\tau=400, Q = 0.03$. }
	\label{DNS channel flow}
\end{figure}

Validity of DNS is checked in Fig. \ref{DNS channel flow}. 
The mean flow has a wall-law region and log-law region (Fig. \ref{DNS channel flow}a).
Figure \ref{DNS channel flow}(b) shows fine vortical structures
visualized by the iso-surface of the
second invariant of the deformation tensor,
which have been observed in previous works \cite{AKM-2001, SH-2002, GLM-2006}.

\subsection{Preparing data for training and test}

\begin{figure}[h]
\begin{center}
  \includegraphics[width=80mm]{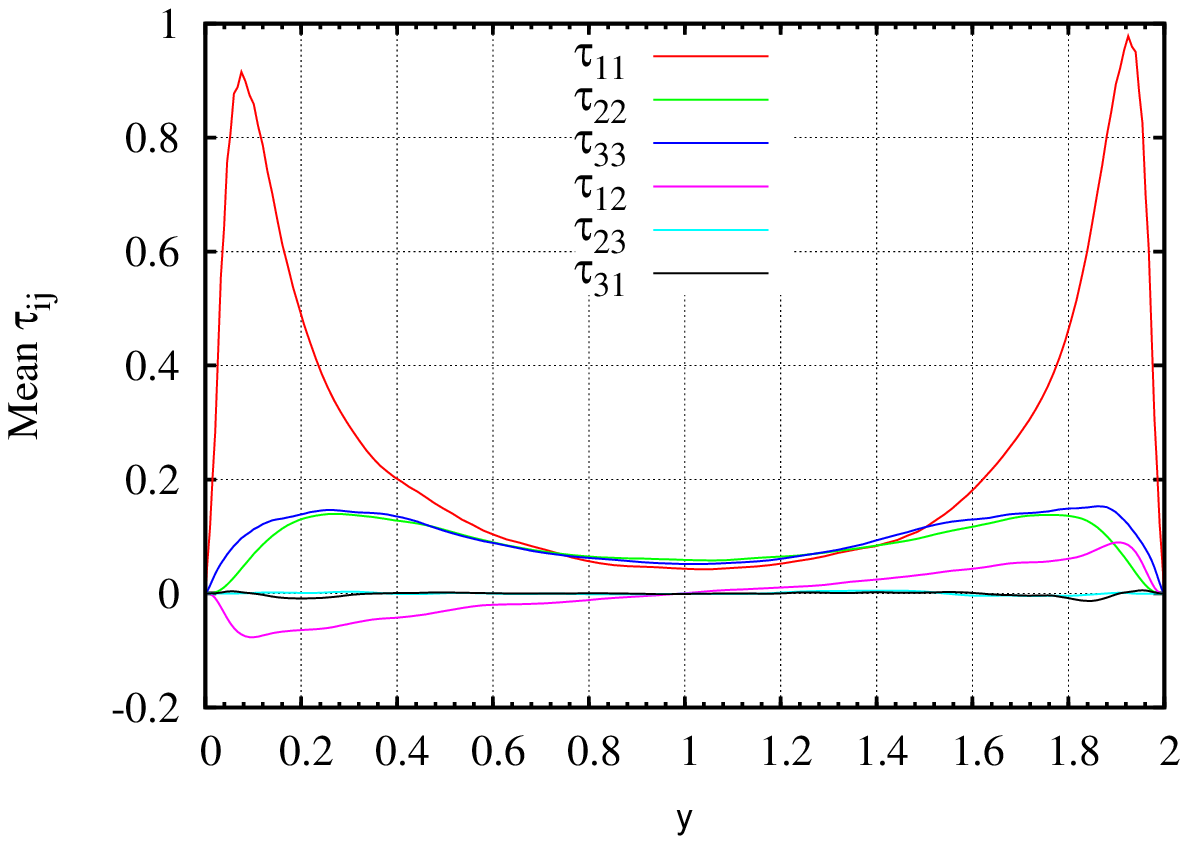}
  \includegraphics[width=80mm]{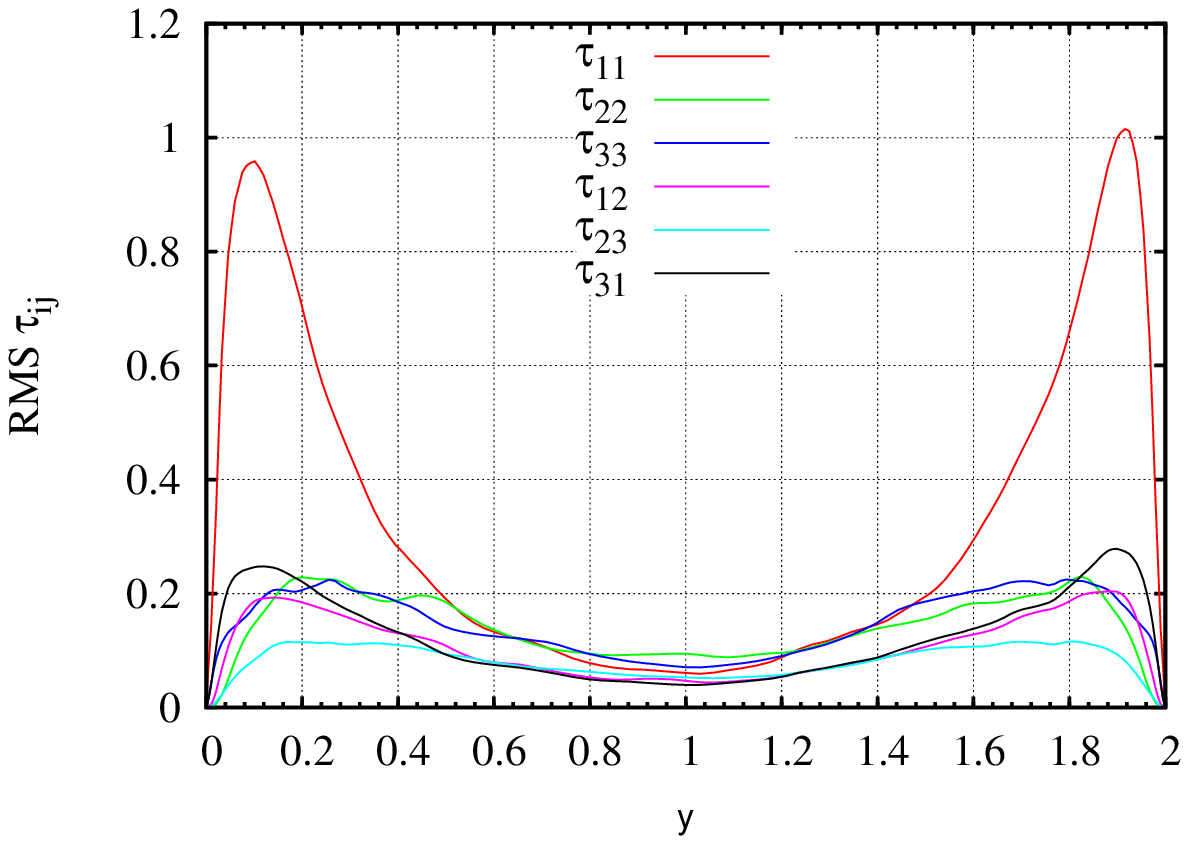}
\caption{Distributions of SGS stress tensor $\tau_{ij}$ averaged in streamwise and spanwise directions ($xz$-plane). 
$Re_\tau = 180, (\overline{\Delta x}^+, \overline{\Delta y}^+_{\rm{max}}, \overline{\Delta z}^+)=(35.3, 9.9, 17.7)$. 
(Left) Average and (right) rms amplitude of fluctuation. 
}
\label{SGSmean-rms}
\end{center}
\end{figure}

The data obtained by DNS are used both for training and for test. 
In training field data obtained by DNS are filtered as normally done in LES
to give the GS flow field on coarse grids, which are used as input variables;
in the present study we use the top-hat filter function. 
The SGS stress is also calculated using the DNS data
and is used as training targets of output variables. 
The values of the filter size and the number of grid points of the coarse grids 
are listed in Table \ref{table filter}; 
for the most part they are set to the underlined parameter values $Re_\tau = 180, (\overline{\Delta x}^+, \overline{\Delta y}^+_{\rm{max}}, \overline{\Delta z}^+)=(35.3, 9.9, 17.7)$. 
The components of SGS stress tensor averaged in the streamwise and spanwise directions
are shown in Fig. \ref{SGSmean-rms}, 
where the rms amplitudes are also shown. 
They are in good agreement with the previous results \cite{KMM1987}. 

The amount of data used for training is rather small; 
typically six positions in the streamwise direction are randomly chosen and 
the data on the corresponding planes parallel to the $yz$ plane are used for training. 
The whole data are used for test of the trained ANN. 

\begin{table}[h]
	\begin{center}
    \vspace{5mm}
		\caption{Parameter values for filtering. Filter size and number of grid points. }
		\def~{\hphantom{0}}
		\begin{tabular}{ccccc}
\hline\hline
			$Re_\tau$ & $\overline{\Delta x}^+$ & $\overline{\Delta y}^{+}_{\rm{max}}$ & $\overline{\Delta z}^{+}$ & $N_x \times N_y \times N_z $\\ \hline
			180&41.7~&15.6~&23.6~&$48\times48\times48$\\
			&43.5~&14.5~&21.7~&$52\times52\times52$\\
			&\underline{35.3}~&\underline{9.9}~&\underline{17.7}~&$\underline{64\times64\times64}$\\
			&23.6~&6.0~&11.8~&$96\times96\times96$\\ \hline

		 	400&39.3~&22.0~&19.6~&$64\times64\times64$\\
		 	&34.3~&17.9~&17.5~&$72\times72\times72$\\
              &29.9~&15.3~&15.0~&$84\times84\times84$\\
              &26.2~&13.3~&13.1~&$96\times96\times96$\\ \hline

			600&29.5~&33.0~&14.7~&$128\times64\times128$\\
			&26.2~&26.9~&13.1~&$144\times72\times144$\\
			&22.4~&23.0~&11.2~&$168\times84\times168$\\
			&19.6~&19.9~&9.8~&$192\times96\times192$\\ \hline

			800&26.2~&10.9~&13.1~&$96\times256\times192$\\
			&19.6~&10.9~&9.8~&$128\times256\times256$\\
			&13.1~&10.9~&6.5~&$192\times256\times384$\\
\hline\hline
  		\end{tabular}
  		\label{table filter}
  	\end{center}
\end{table}

\subsection{Artificial neural network (ANN)}

\begin{figure}[h]
	\centerline{\includegraphics[width=40mm]{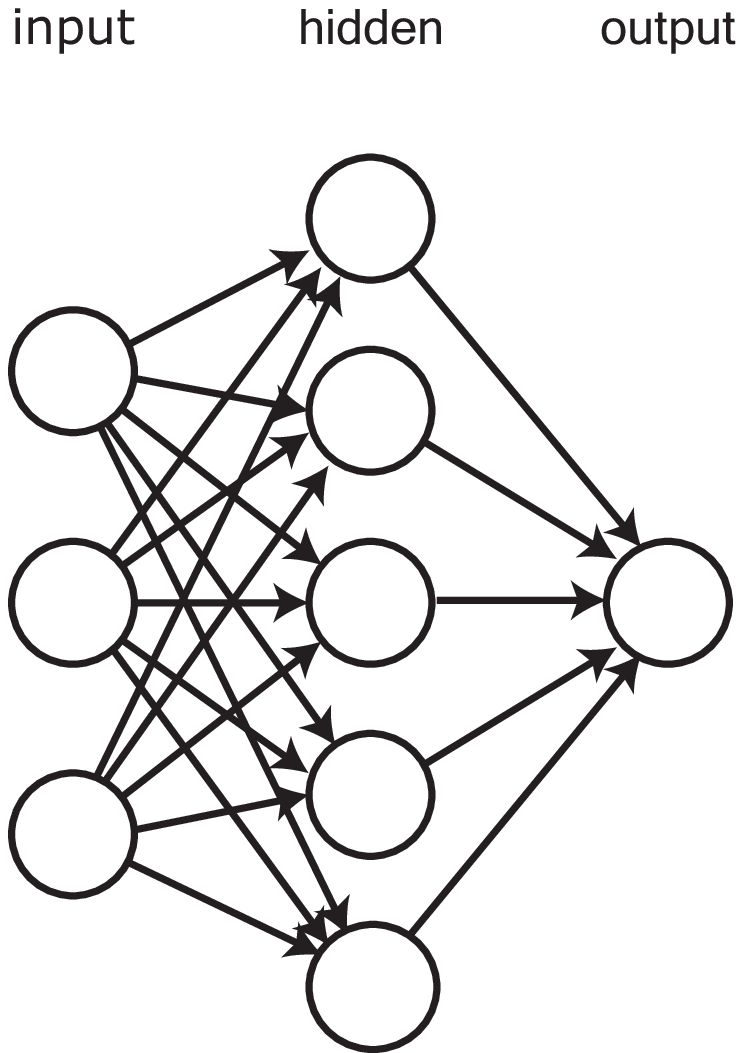}
\hspace{1cm}
\includegraphics[width=60mm]{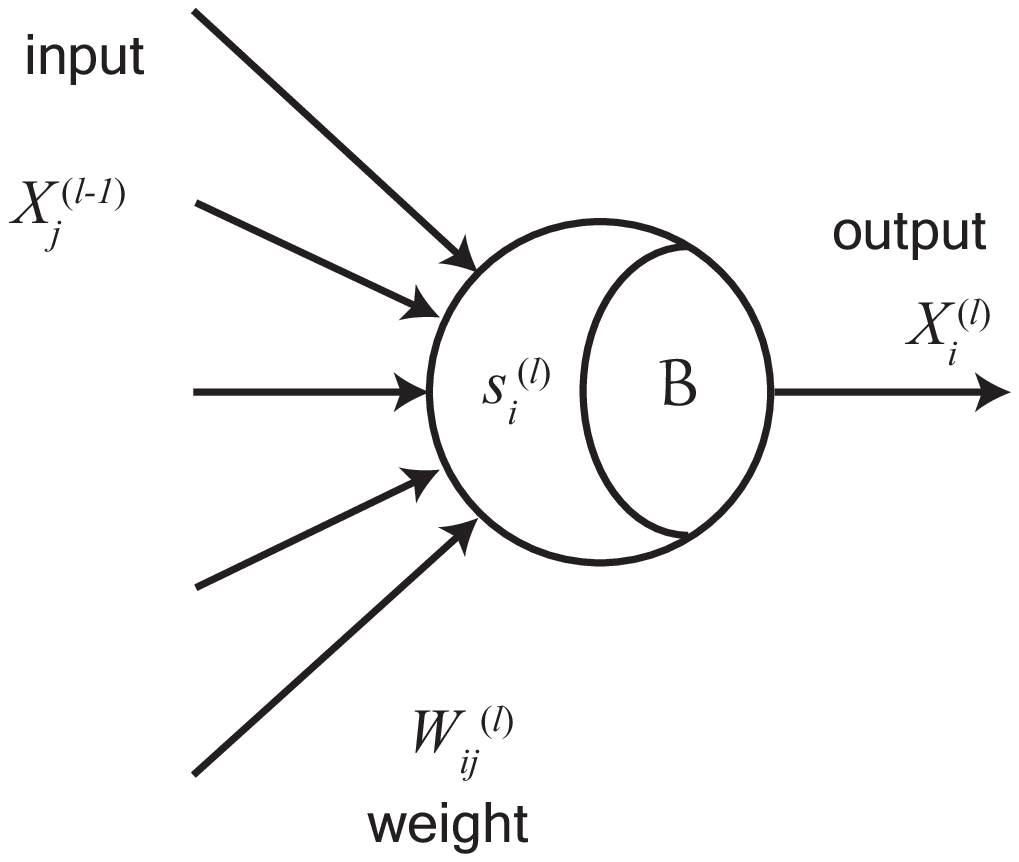}
}   
  \vspace{5mm}
	\caption{Schematic diagram of artificial neural network (ANN). 
(Left) Network structure; 
(right) single neuron. }
	\label{Neural network}
\end{figure}

A feedforward neural network is employed in establishing a functional relation 
between the GS flow field and the SGS stress tensor. 
Figure \ref{Neural network} shows the schematic diagram of ANN 
and a single neuron. 
Let us consider ANN which consists of $L$ layers. 
A single neuron of the $l$-th layer receives a set of inputs $\left\{X_j^{(l-1)}\right\}$ 
and then outputs $X_i^{(l)}$ which is calculated as 
\begin{eqnarray}
X_i^{(l)} &=& {\cal{B}}\left(s_i^{(l)}+b_i^{(l)}\right), \\
s_i^{(l)} &=& \sum_j W_{ij}^{(l)} X_j^{(l-1)},  
\end{eqnarray}
where $\cal{B}$ is the activation function, 
$b_i^{(l)}$ is the bias parameter, and $W_{ij}^{(l)}$ is the weight. 
The bias parameters and the weights are corrected iteratively 
so that the final output $X^{(L)}$ approximates well the given SGS stress. 
The data of the first layer $\left\{X_j^{(1)}\right\}$ are given by the GS flow field. 

In the present study ANN consists of three layers: 
the input, hidden, and output layers. 
For simplicity, each independent component of the SGS stress tensor 
is dealt with separately; 
in other words, six ANNs are trained to approximate all components of the SGS stress tensor. 
This point will be discussed from the viewpoint of symmetry 
in the final section. 
The activation function is the sigmoid function ${\cal{B}}(z) = 1/(1+e^{-\alpha z})$. 
The back propagation is used as a method for training which 
optimizes the bias parameters and weights iteratively to 
minimize the difference between the output and the given SGS stress 
$\sum |X^{(L)}-\tau_{ij}|^2$. 
The number of neurons of the input layer depends on the choice of input variables 
as described below, while the output layer consists of a single neuron. 
The number of neurons of the hidden layer is $100$ unless stated explicitly, 
while the dependence on it is checked in Sec.~\ref{sec-res-basic}. 

We should be careful in choosing the input variables 
since it is important for effective and successful learning. 
Although there are many possibilities for the combination of input variables, 
we limit ourselves to pointwise correspondence between the input and output variables; 
namely, in order to approximate the SGS stress at one point 
the GS data at the same point are used as input variables. 
We test four sets of input variables:  
(i) $\{\mathbd{S}, y\}$;  
(ii) $\{\mathbd{S}, \mathbd{\Omega}, y\}$; 
(iii) $\{\mathbd{\nabla} \overline{\mathbd{u}}, y\}$; 
(iv) $\{\mathbd{\nabla} \overline{\mathbd{u}}\}$,  
where $\mathbd{S}=[{\mathbd{\nabla}}\overline{\mathbd{u}}+({\mathbd{\nabla}}\overline{\mathbd{u}})^T]/2$ and 
$\mathbd{\Omega}=[{\mathbd{\nabla}}\overline{\mathbd{u}}-({\mathbd{\nabla}}\overline{\mathbd{u}})^T]/2$.  
The first set can give the Smagorinsky model 
$\tau_{ij} = -2C_S \overline{\Delta}^2 (2S_{kl}S_{kl})^{1/2} S_{ij} +(1/3)\tau_{kk}\delta_{ij}$, 
while the position $y$ is included to take account of possible dependence on 
the wall-normal direction. 
The GS vorticity is added in the second set 
since it improves the accuracy of the SGS model in some flows \cite{K-2005}. 
The third set is equivalent to the second one 
in the sense that $\mathbd{\nabla} \overline{\mathbd{u}}=\mathbd{S}+\mathbd{\Omega}$; 
in practice, however, 
the results of learning can be different. 
The last set is included to check the role of the position $y$. 

\section{Results}

\subsection{Choice of input variables}

\begin{figure}[h]
\begin{center}
{\includegraphics[width=80mm]{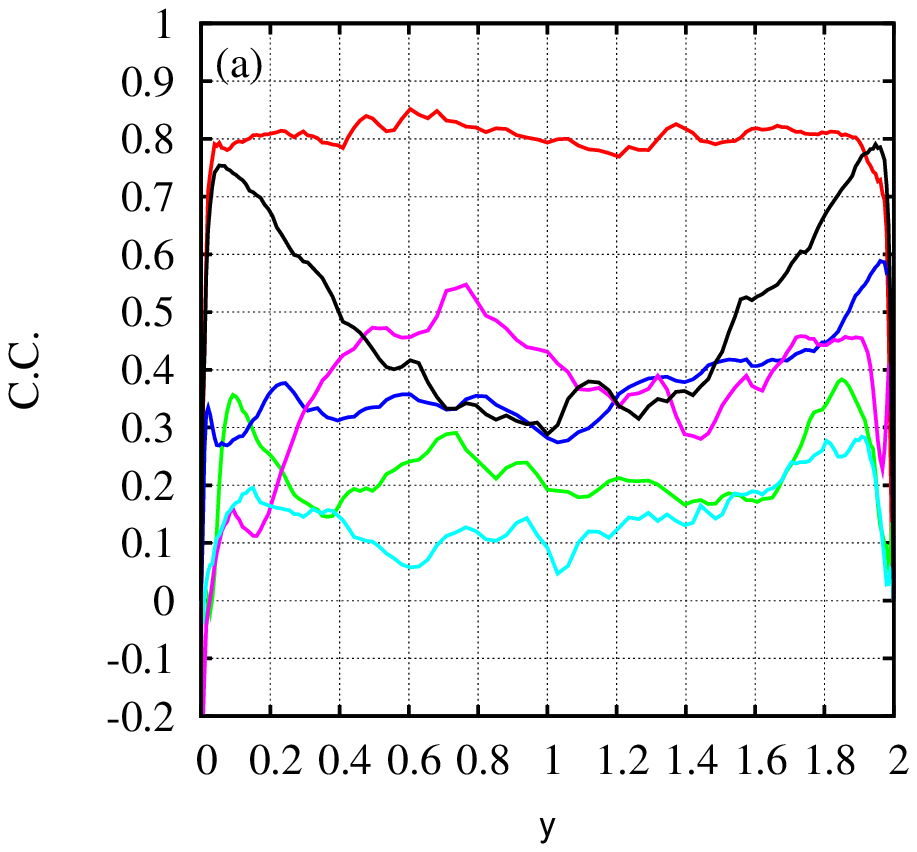}}
{\includegraphics[width=80mm]{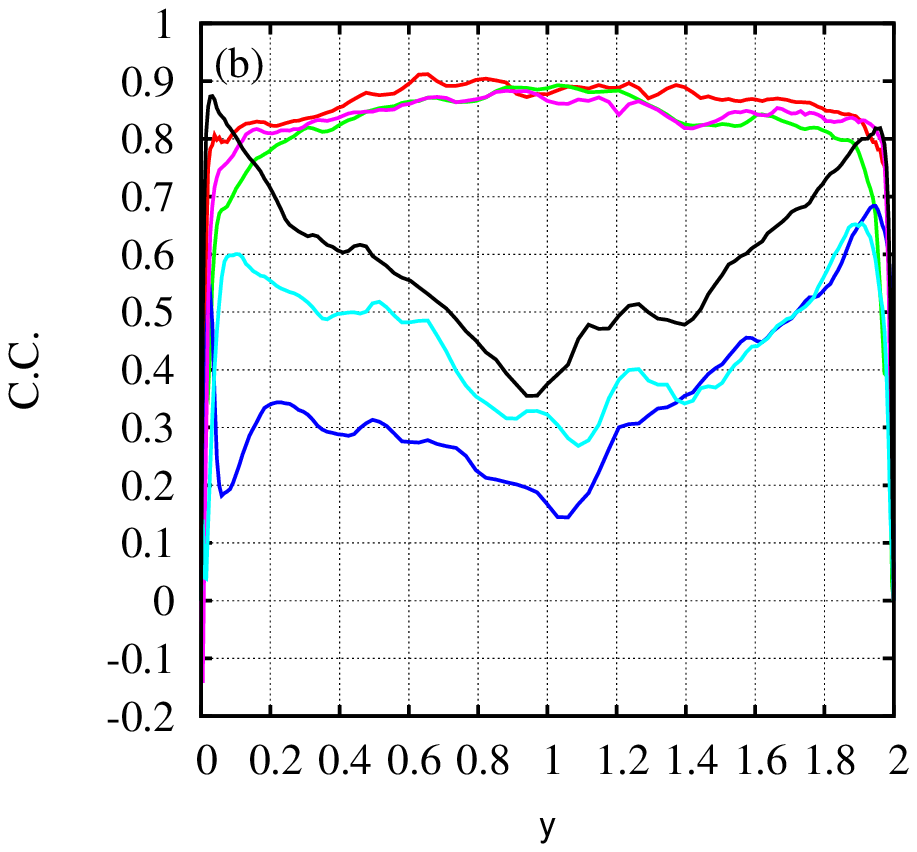}}
{\includegraphics[width=80mm]{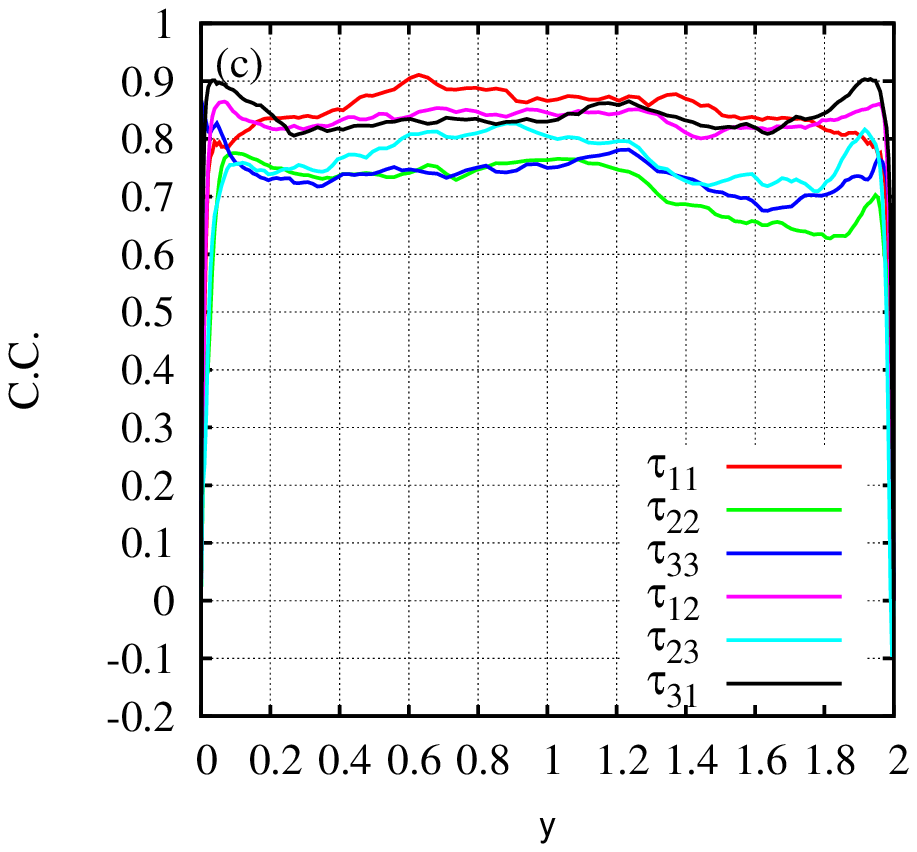}}
{\includegraphics[width=80mm]{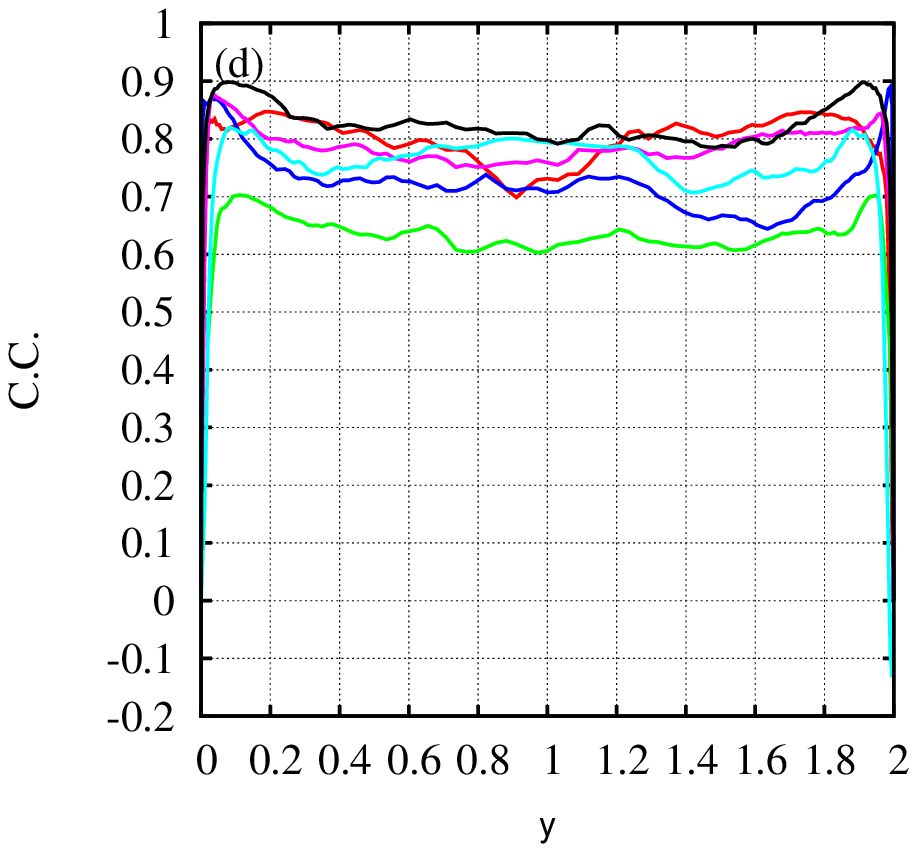}}
\caption{Correlation coefficients between the SGS stress $\tau_{ij}^{(DNS)}$ calculated using DNS data 
and $\tau_{ij}^{(ANN)}$ predicted by trained ANN. 
Correlation coefficients are averaged in streamwise and spanwise directions ($xz$-plane). 
$Re_\tau = 180, (\overline{\Delta x}^+, \overline{\Delta y}^+_{\rm{max}}, \overline{\Delta z}^+)=(35.3, 9.9, 17.7)$. 
(a) $\{\mathbd{S}, y\}$, 
(b) $\{\mathbd{S}, \mathbd{\Omega}, y\}$, 
(c) $\{\mathbd{\nabla} \overline{\mathbd{u}}, y\}$, 
(d) $\{\mathbd{\nabla} \overline{\mathbd{u}}\}$. 
}
\label{fig:all-CC}
\end{center}
\end{figure}

First, we investigate which set of input variables results in successful learning. 
Figure \ref{fig:all-CC} shows correlation coefficients
between the SGS stress $\tau_{ij}^{(DNS)}$ calculated using DNS data 
and $\tau_{ij}^{(ANN)}$ predicted by trained ANN for the four sets of input variables 
for $Re_\tau = 180, (\overline{\Delta x}^+, \overline{\Delta y}^+_{\rm{max}}, \overline{\Delta z}^+)=(35.3, 9.9, 17.7)$.   
The correlation coefficients are calculated by integrating in $x$ and $z$ directions 
\begin{eqnarray*}
{\rm{C.C.}}(y) &=& \frac{\avrgxz{\left(\tau_{ij}^{(DNS)} - \avrgxz{\tau_{ij}^{(DNS)}}\right)\left(\tau_{ij}^{(ANN)} - \avrgxz{\tau_{ij}^{(ANN)}}\right)}}{\left[\avrgxz{\left(\tau_{ij}^{(DNS)}-\avrgxz{\tau_{ij}^{(DNS)}}\right)^2}\right]^{1/2}\left[\avrgxz{\left(\tau_{ij}^{(ANN)}-\avrgxz{\tau_{ij}^{(ANN)}}\right)^2}\right]^{1/2}}, \\ 
\avrgxz{f} &=& \frac{1}{L_xL_z}\int_0^{L_x} dx \int_0^{L_z} dz \ f(x,y,z). 
\end{eqnarray*}
High correlation implies successful learning. 
It is seen that learning is most successful 
when $\{\mathbd{\nabla} \overline{\mathbd{u}}, y\}$ is used as input variables 
as the averaged correlation coefficient 
$\overline{{\rm{C.C.}}} = (1/L_y) \int_0^{L_y} dy \ {\rm{C.C.}}(y)$ 
exceeds $0.7$ for all six components as shown in Table \ref{CC-input-table1}, 
although $\{\mathbd{\nabla} \overline{\mathbd{u}}\}$ gives comparable success; 
the difference of the correlation coefficients between the two sets is $0 \sim 4\%$. 
Thus the distance from the wall is not important very much. 
The Smagorinsky type $\{\mathbd{S}, y\}$ gives poor results 
as the averaged correlation coefficient exceeds $0.7$ for only one component $\tau_{11}$;   
the correlation coefficients of the other components are small in the entire region 
besides that it is large for $\tau_{31}$ near the wall. 
This result is reasonable since it is known that 
the eddy-viscosity type model has little correlation with the real SGS stress \cite{LMK-1994}. 
There is a little improvement by including $\mathbd{\Omega}$ 
as the set $\{\mathbd{S}, \mathbd{\Omega}, y\}$ gives three correlation coefficients which exceed $0.7$; 
the correlation coefficients of $\tau_{33}$, $\tau_{23}$, and $\tau_{31}$ are small 
in the central region where both the average and rms amplitude of these components are small.   
However, it should be pointed out that
the set $\{\mathbd{S}, \mathbd{\Omega}, y\}$ 
is not good as $\{\mathbd{\nabla} \overline{\mathbd{u}}, y\}$, 
although the two sets have essentially the same degree of freedom.   

Among the components of the SGS stress tensor 
$\tau_{11}$ is the easiest component for regression 
as all four sets give correlation coefficients larger than $0.7$; 
this is because $\tau_{11}$ is largest in magnitude (Fig. \ref{SGSmean-rms}). 
On the other hand, $\tau_{23}$ and $\tau_{31}$, 
of which averages are zero and amplitudes are small (Fig. \ref{SGSmean-rms}), 
are difficult to be approximated 
as the correlation coefficients are small 
for both $\{\mathbd{S}, y\}$ and $\{\mathbd{S}, \mathbd{\Omega}, y\}$. 
In the rest of the paper 
the set of input variables is fixed to $\{\mathbd{\nabla} \overline{\mathbd{u}}, y\}$.  

\begin{table}[h]
 	\begin{center}
    \vspace{5mm}
		\caption{Correlation coefficients between the SGS stress $\tau_{ij}^{(DNS)}$ calculated using DNS data 
and $\tau_{ij}^{(ANN)}$ predicted by trained ANN. 
Correlation coefficients are averaged in the whole domain. }
		\def~{\hphantom{0}}
  		\begin{tabular}{ccccccc}
                  \hline\hline
			Input variables &$\tau_{11}$&$\tau_{22}$&$\tau_{33}$&$\tau_{12}$&$\tau_{23}$&$\tau_{31}$\\ \hline
		 	$\{\mathbd{S}, y\}$ &0.793~&0.292~&0.390~&0.320~&0.148~&0.489~\\
			$\{\mathbd{S}, \mathbd{\Omega}, y\}$ &0.776~&0.769~&0.344~&0.773~&0.368~&0.524~\\
			$\{\mathbd{\nabla} \overline{\mathbd{u}}, y\}$ &0.804~&0.713~&0.728~&0.791~&0.730~&0.821~\\
			$\{\mathbd{\nabla} \overline{\mathbd{u}}\}$ &0.767~&0.670~&0.710~&0.776~&0.720~&0.820~\\
                  \hline\hline
  		\end{tabular}
  		\label{CC-input-table1}
  	\end{center}
\end{table}

\subsection{How successful is the learning?}

\begin{figure}[h]
\begin{minipage}{0.45\hsize}
\begin{flushleft}
\hspace*{15mm}(a) $\tau_{11}^{(DNS)}$
\end{flushleft}
\vspace{-5mm}
   \centering
{\includegraphics[width=80mm]{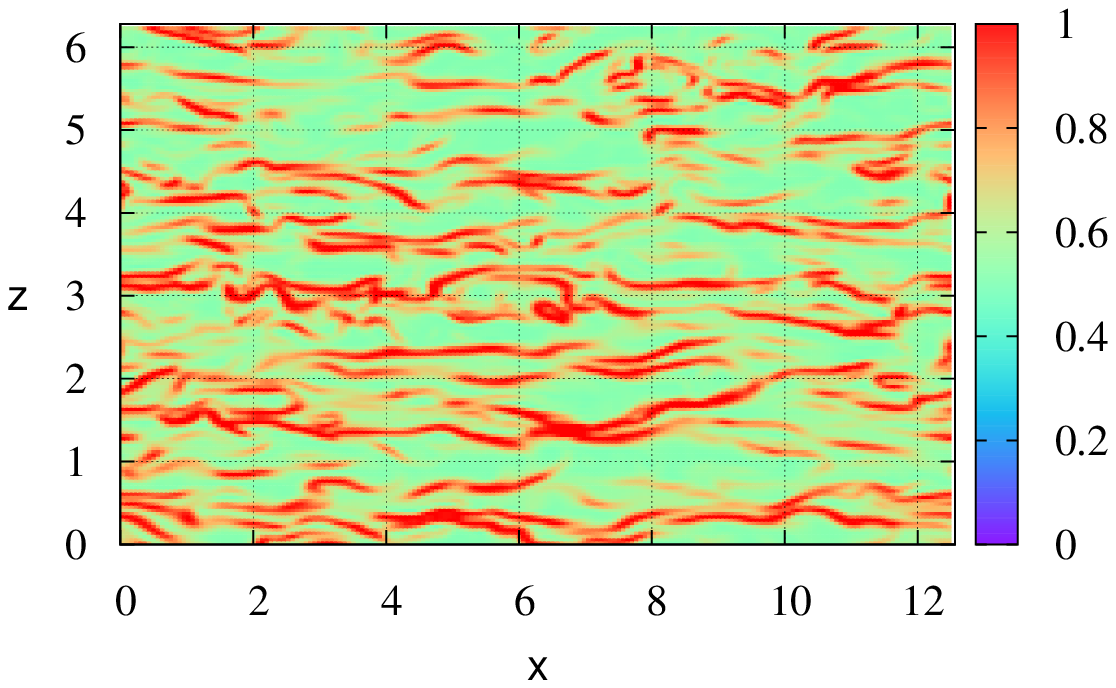}}
\end{minipage}
\hspace{5mm}
\begin{minipage}{0.45\hsize}
\begin{flushleft}
\hspace*{15mm}(b) $\tau_{11}^{(ANN)}$
\end{flushleft}
\vspace{-5mm}
   \centering
{\includegraphics[width=80mm]{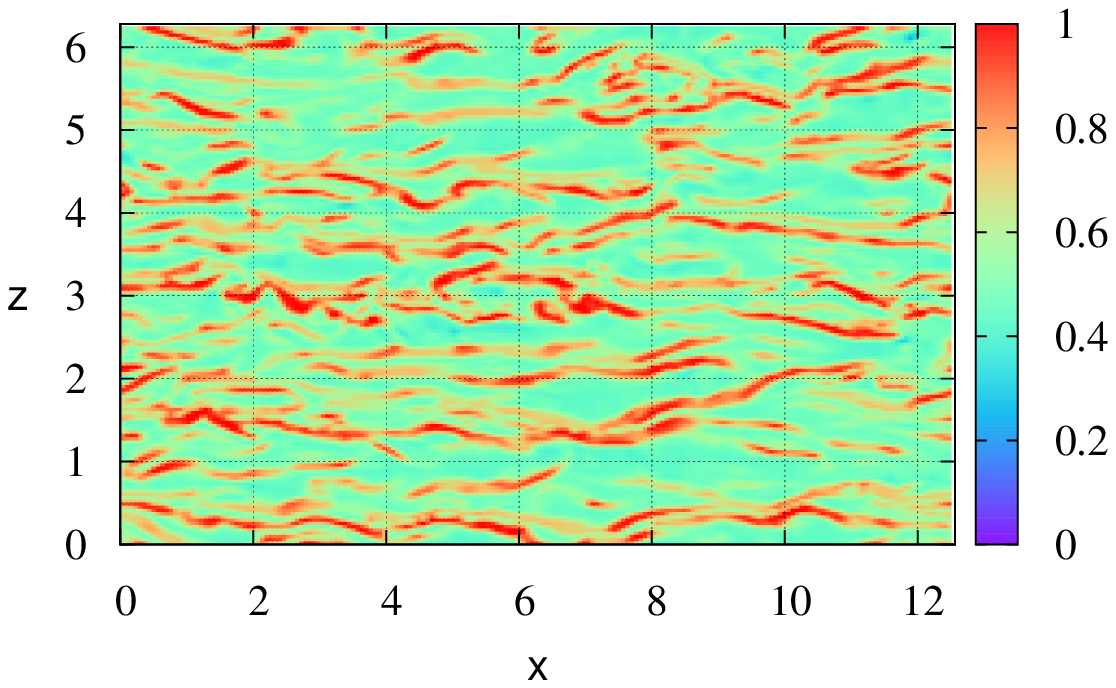}}
\end{minipage}
%
%
\begin{minipage}{0.45\hsize}
\begin{flushleft}
\hspace*{15mm}(c) $\tau_{22}^{(DNS)}$
\end{flushleft}
\vspace{-5mm}
   \centering
{\includegraphics[width=80mm]{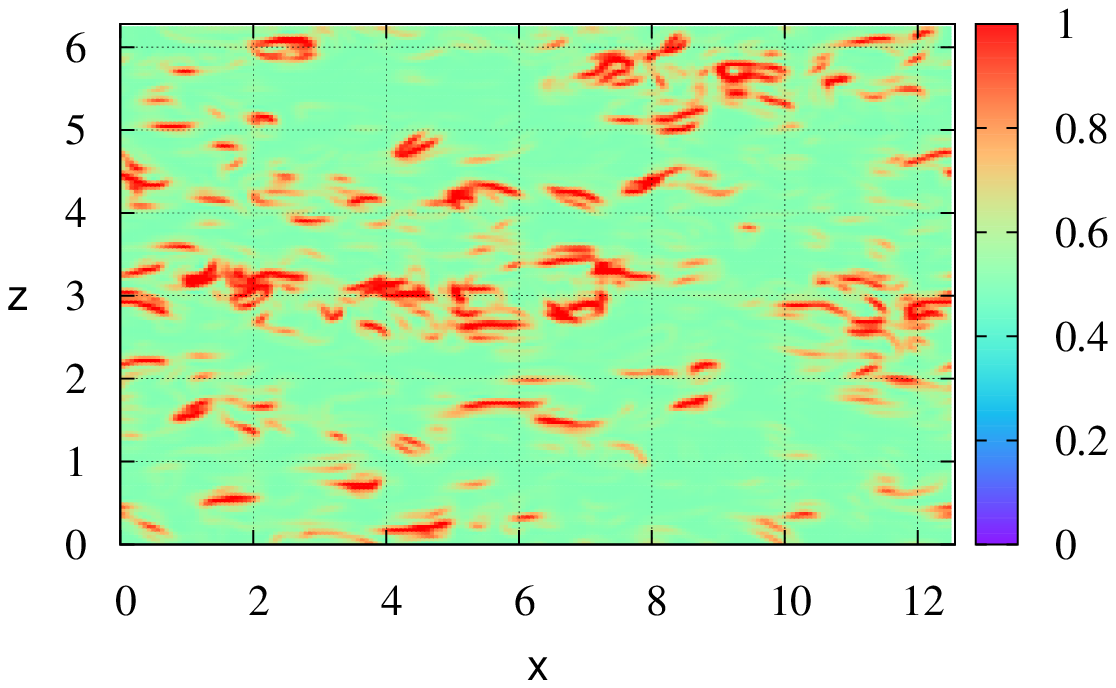}}
\end{minipage}
\hspace{5mm}
\begin{minipage}{0.45\hsize}
\begin{flushleft}
\hspace*{15mm}(d) $\tau_{22}^{(ANN)}$
\end{flushleft}
\vspace{-5mm}
   \centering
{\includegraphics[width=80mm]{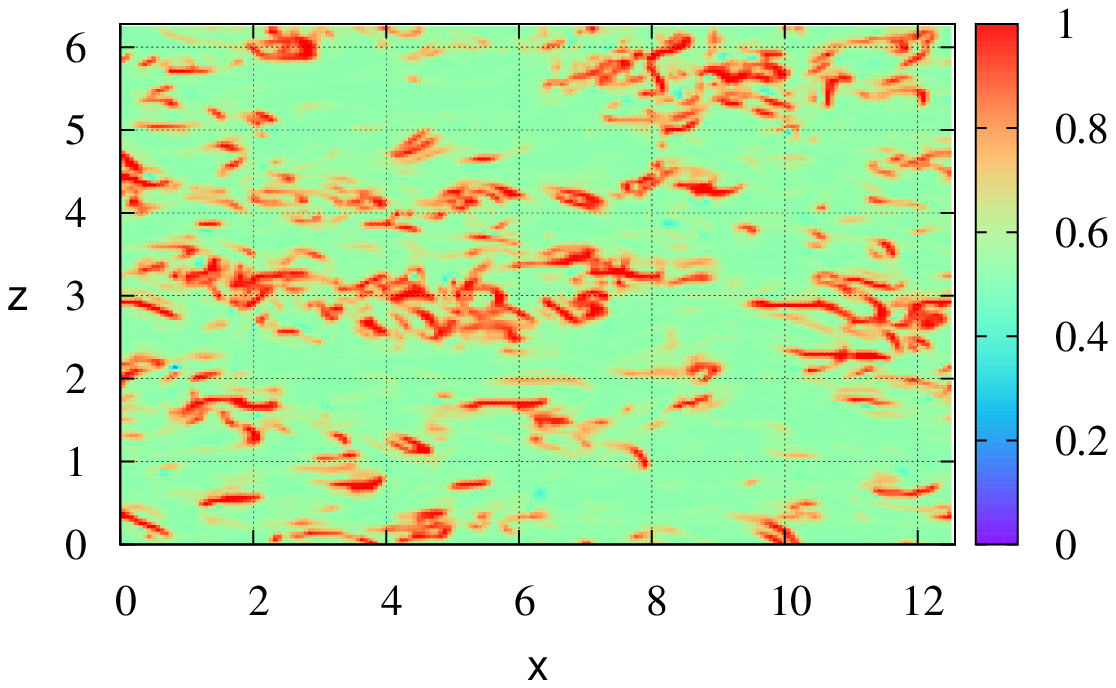}}
\end{minipage}
%
%
\begin{minipage}{0.45\hsize}
\begin{flushleft}
\hspace*{15mm}(e) $\tau_{33}^{(DNS)}$
\end{flushleft}
\vspace{-5mm}
   \centering
{\includegraphics[width=80mm]{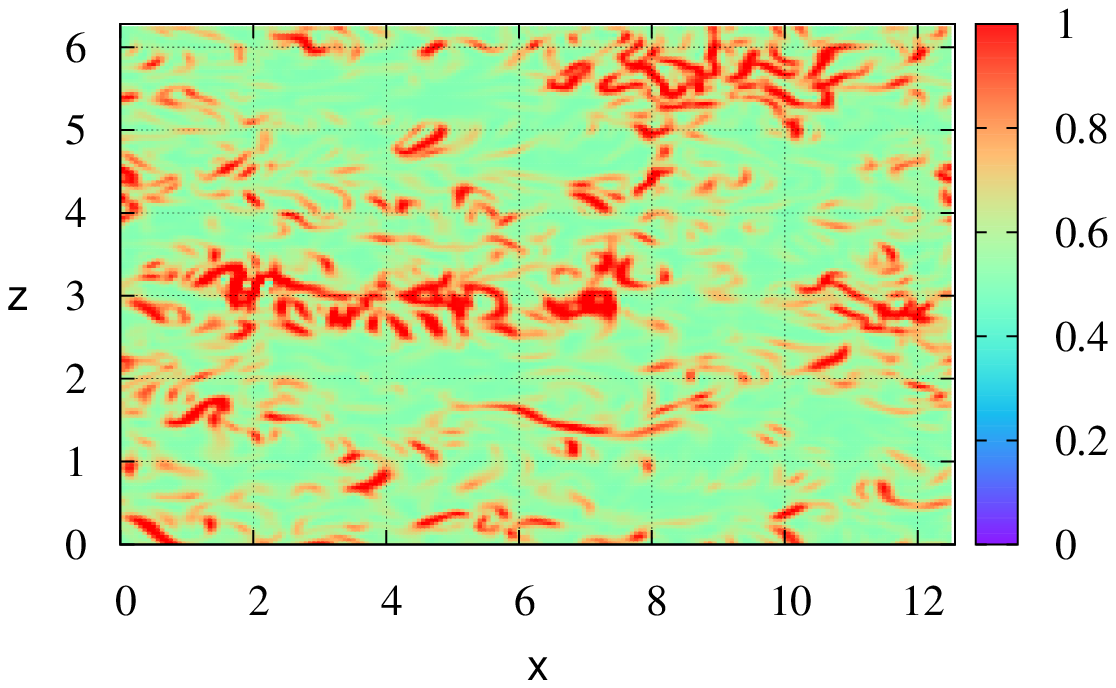}}
\end{minipage}
\hspace{5mm}
\begin{minipage}{0.45\hsize}
\begin{flushleft}
\hspace*{15mm}(f) $\tau_{33}^{(ANN)}$
\end{flushleft}
\vspace{-5mm}
   \centering
{\includegraphics[width=80mm]{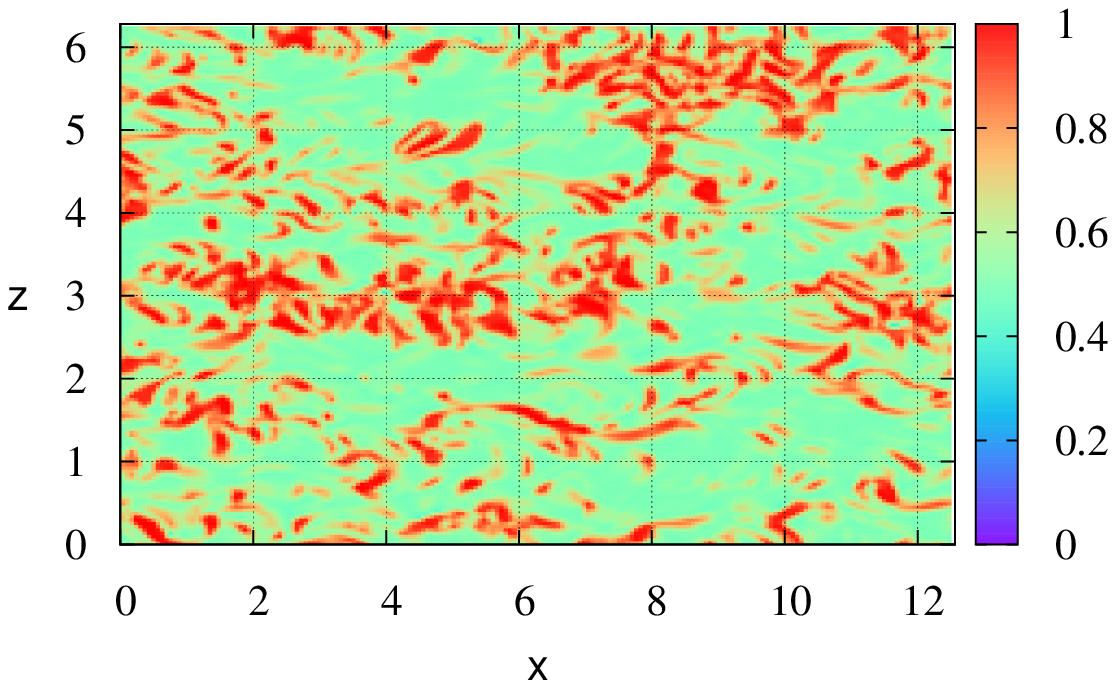}}
\end{minipage}
\caption{Spatial distributions of SGS stress tensor at $y = 0.1$. 
Comparison between $\tau_{ij}^{(DNS)}$ and $\tau_{ij}^{(ANN)}$. 
Diagonal components. 
$Re_\tau = 180, (\overline{\Delta x}^+, \overline{\Delta y}^+_{\rm{max}}, \overline{\Delta z}^+)=(35.3, 9.9, 17.7)$. }
\label{fig:Dis}
\end{figure}

\begin{figure}[h]
\begin{minipage}{0.45\hsize}
\begin{flushleft}
\hspace*{15mm}(a) $\tau_{12}^{(DNS)}$
\end{flushleft}
\vspace{-5mm}
   \centering
{\includegraphics[width=80mm]{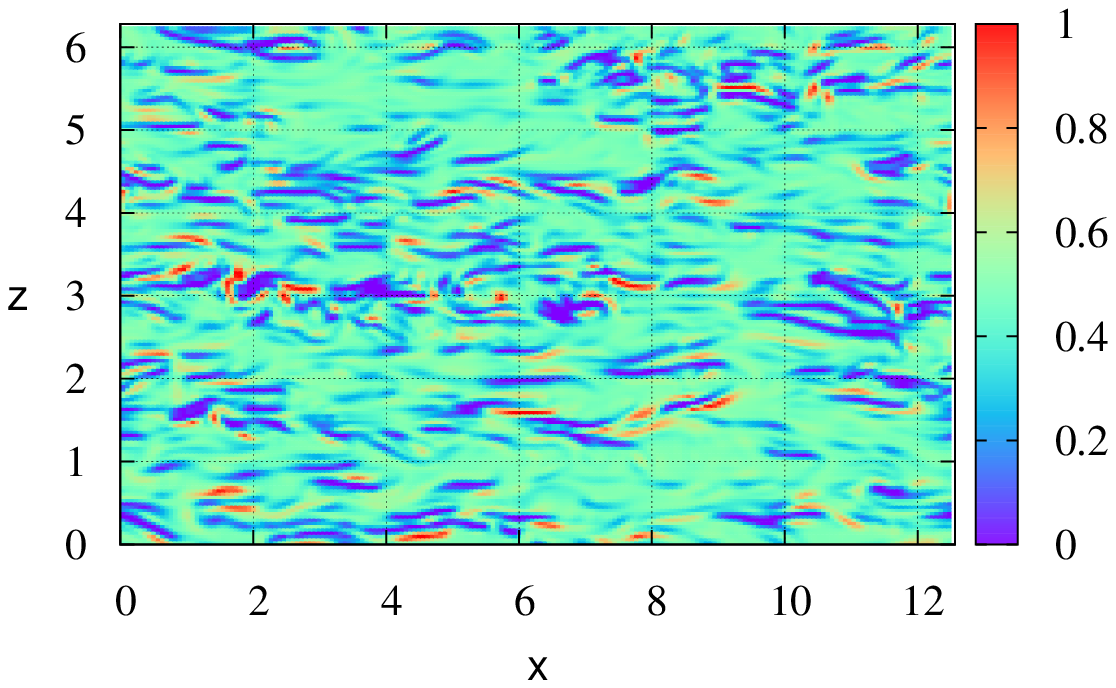}}
\end{minipage}
\hspace{5mm}
\begin{minipage}{0.45\hsize}
\begin{flushleft}
\hspace*{15mm}(b) $\tau_{12}^{(ANN)}$
\end{flushleft}
\vspace{-5mm}
   \centering
{\includegraphics[width=80mm]{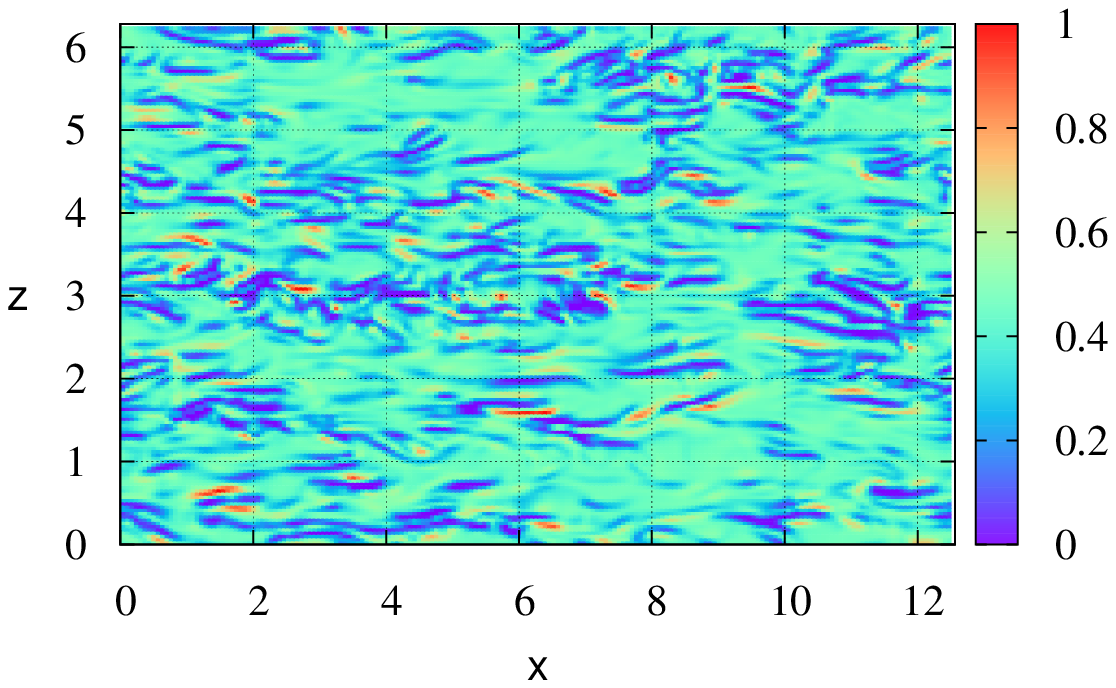}}
\end{minipage}
\begin{minipage}{0.45\hsize}
\begin{flushleft}
\hspace*{15mm}(c) $\tau_{23}^{(DNS)}$
\end{flushleft}
\vspace{-5mm}
   \centering
{\includegraphics[width=80mm]{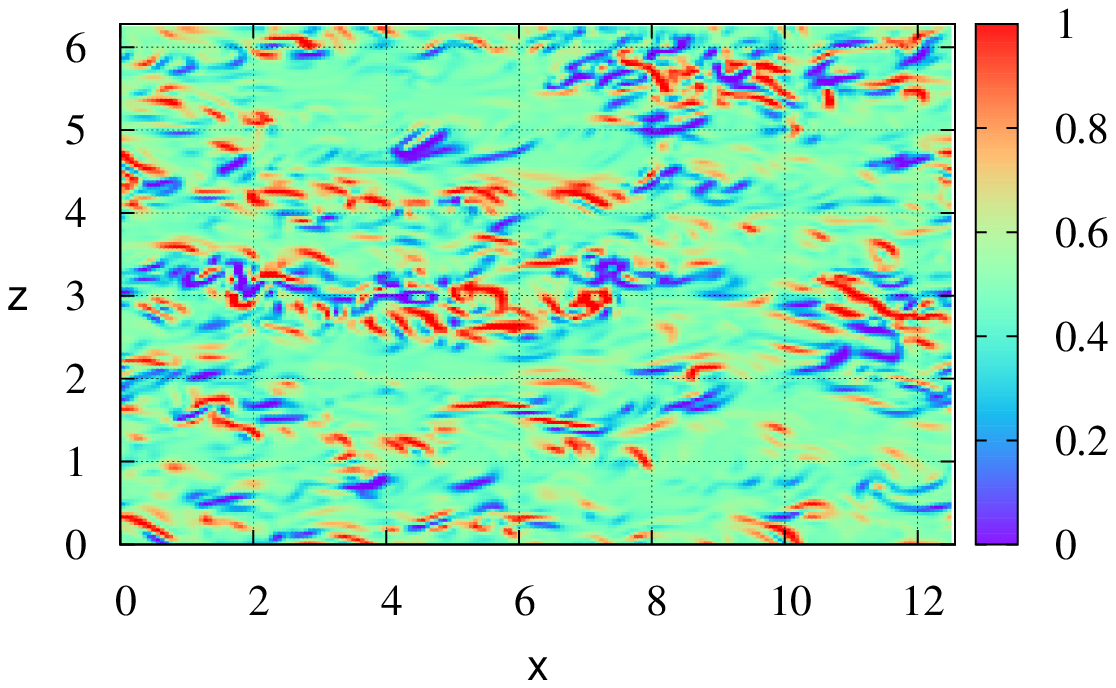}}
\end{minipage}
\hspace{5mm}
\begin{minipage}{0.45\hsize}
\begin{flushleft}
\hspace*{15mm}(d) $\tau_{23}^{(ANN)}$
\end{flushleft}
\vspace{-5mm}
   \centering
{\includegraphics[width=80mm]{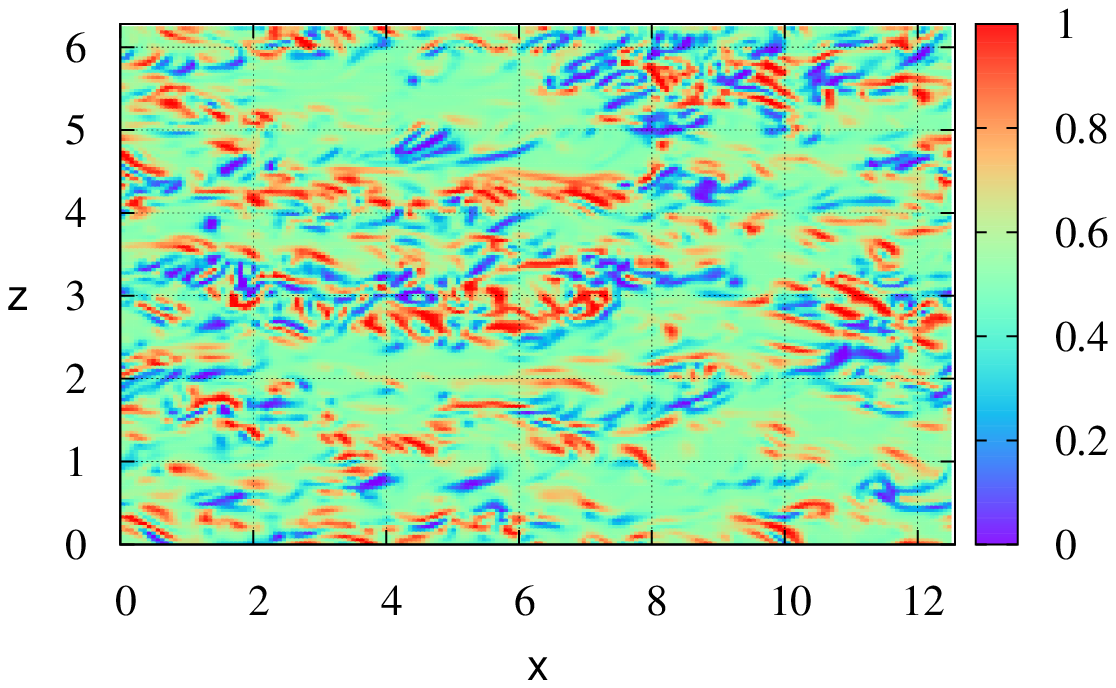}}
\end{minipage}
\begin{minipage}{0.45\hsize}
\begin{flushleft}
\hspace*{15mm}(e) $\tau_{31}^{(DNS)}$
\end{flushleft}
\vspace{-5mm}
   \centering
{\includegraphics[width=80mm]{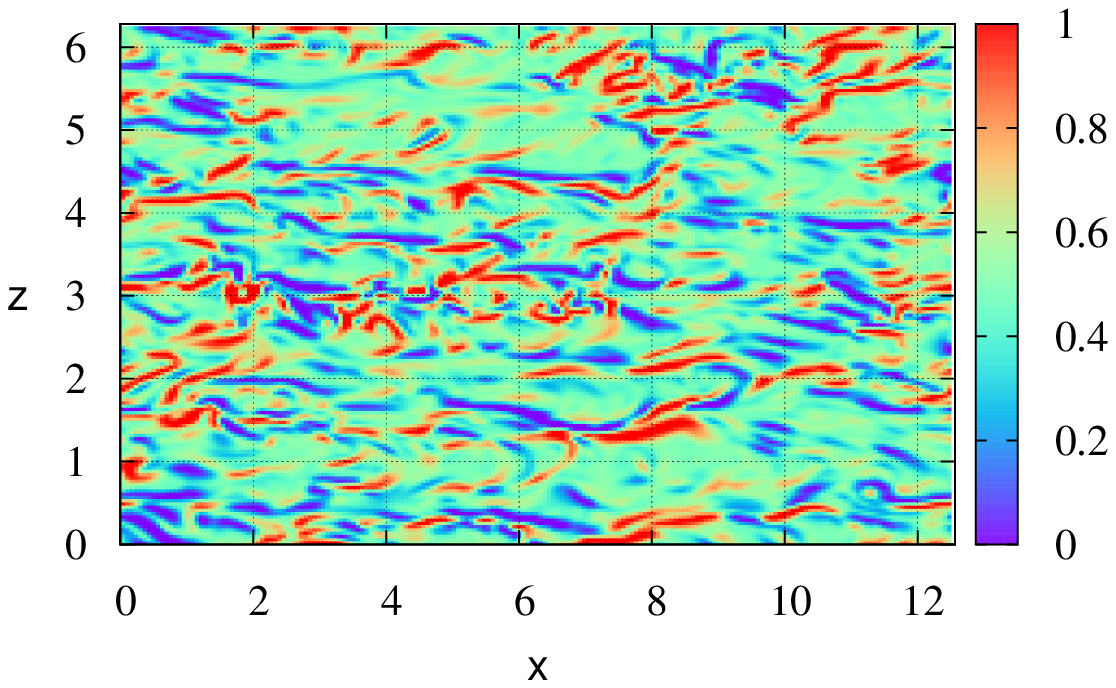}}
\end{minipage}
\hspace{5mm}
\begin{minipage}{0.45\hsize}
\begin{flushleft}
\hspace*{15mm}(f) $\tau_{31}^{(ANN)}$
\end{flushleft}
\vspace{-5mm}
   \centering
{\includegraphics[width=80mm]{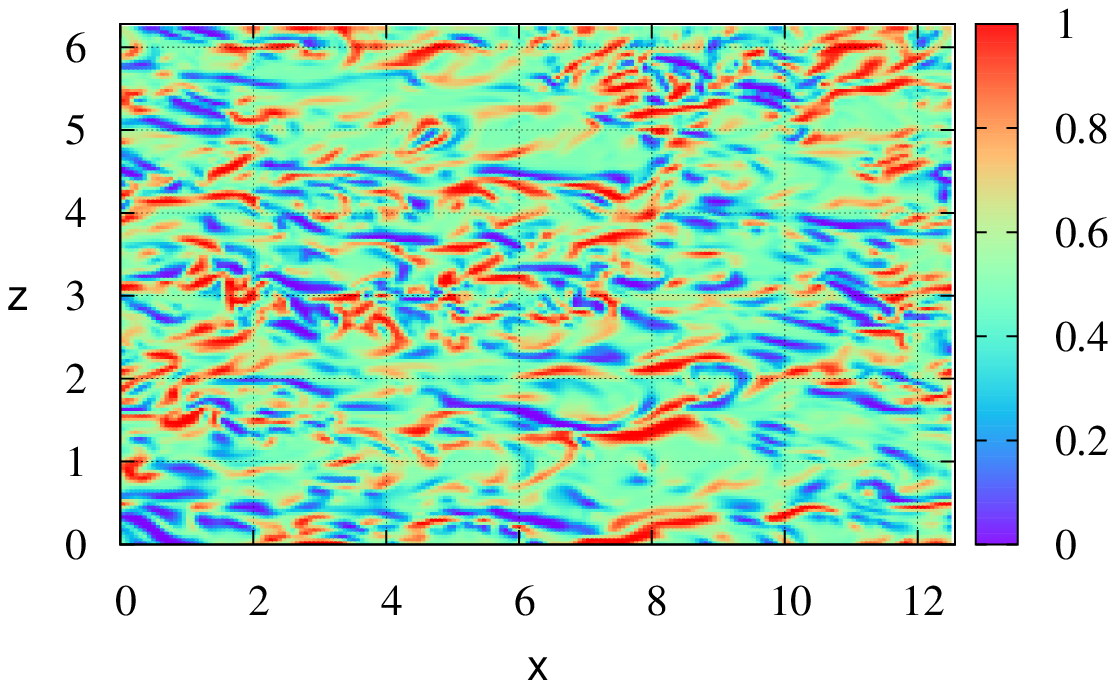}}
\end{minipage}
\caption{
Spatial distributions of SGS stress tensor at $y = 0.1$. 
Same as in Fig. \ref{fig:Dis} but for off-diagonal components. }
\label{fig:Dis2}
\end{figure}

\begin{figure}[h]
\begin{minipage}{0.45\hsize}
\begin{flushleft}
\hspace*{15mm}(a) $\tau_{11}$
\end{flushleft}
\vspace{-5mm}
   \centering
{\includegraphics[width=80mm]{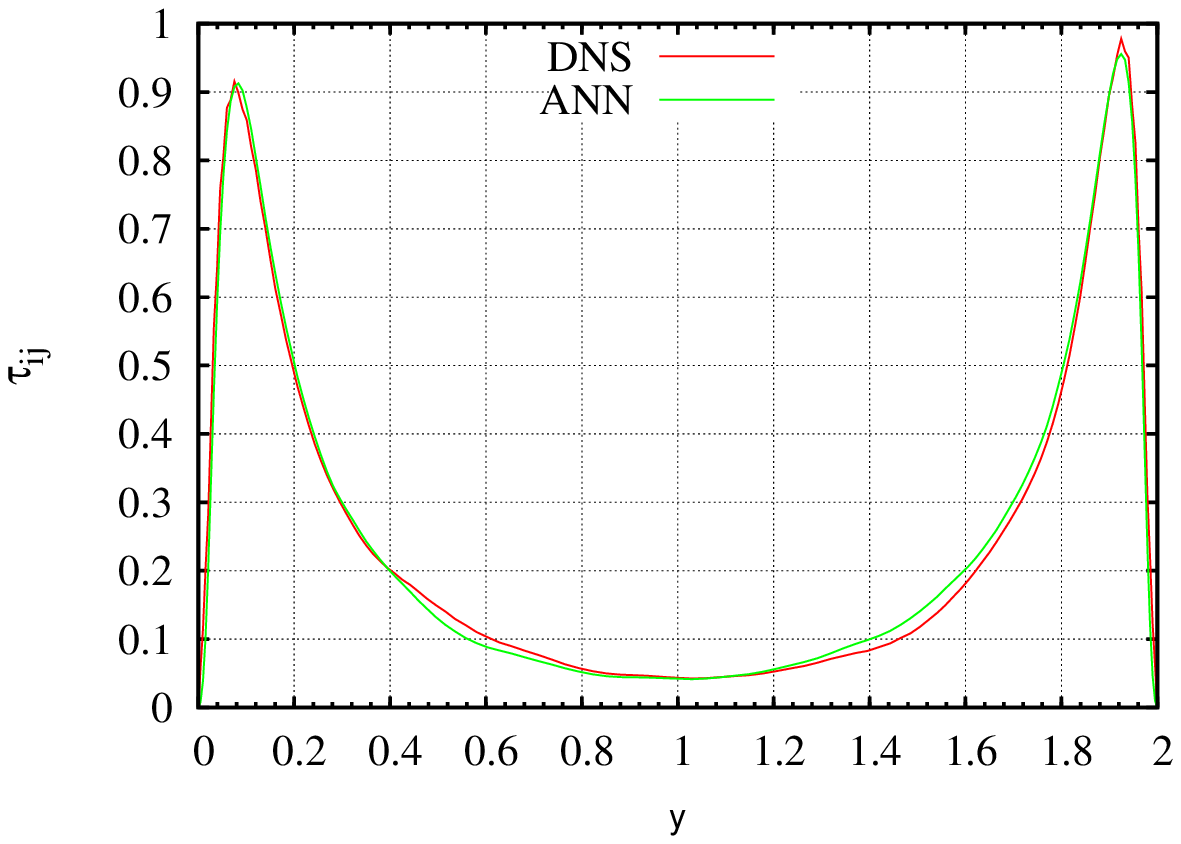}}
\end{minipage}
\hspace{5mm}
\begin{minipage}{0.45\hsize}
\begin{flushleft}
\hspace*{15mm}(b) $\tau_{22}$
\end{flushleft}
\vspace{-5mm}
   \centering
{\includegraphics[width=80mm]{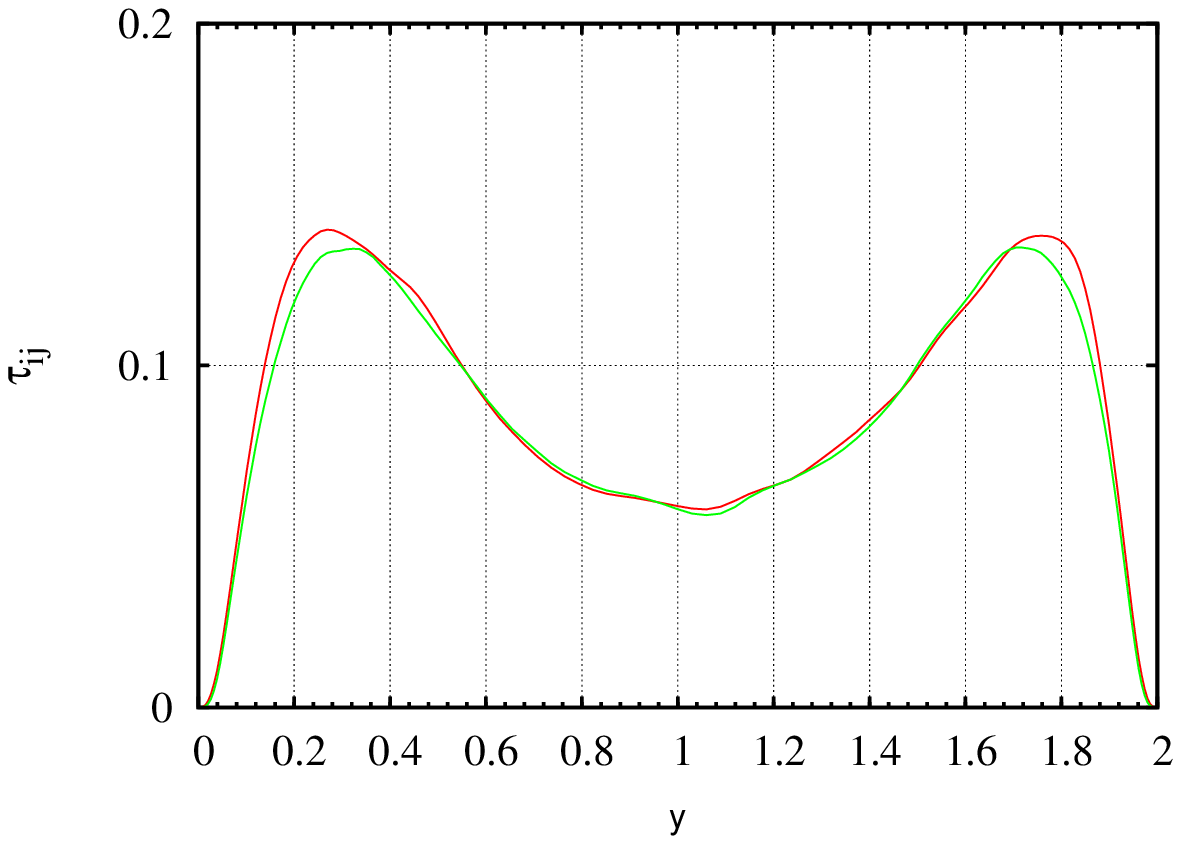}}
\end{minipage}
\vspace{5mm}

\begin{minipage}{0.45\hsize}
\begin{flushleft}
\hspace*{15mm}(c) $\tau_{33}$
\end{flushleft}
\vspace{-5mm}
   \centering
{\includegraphics[width=80mm]{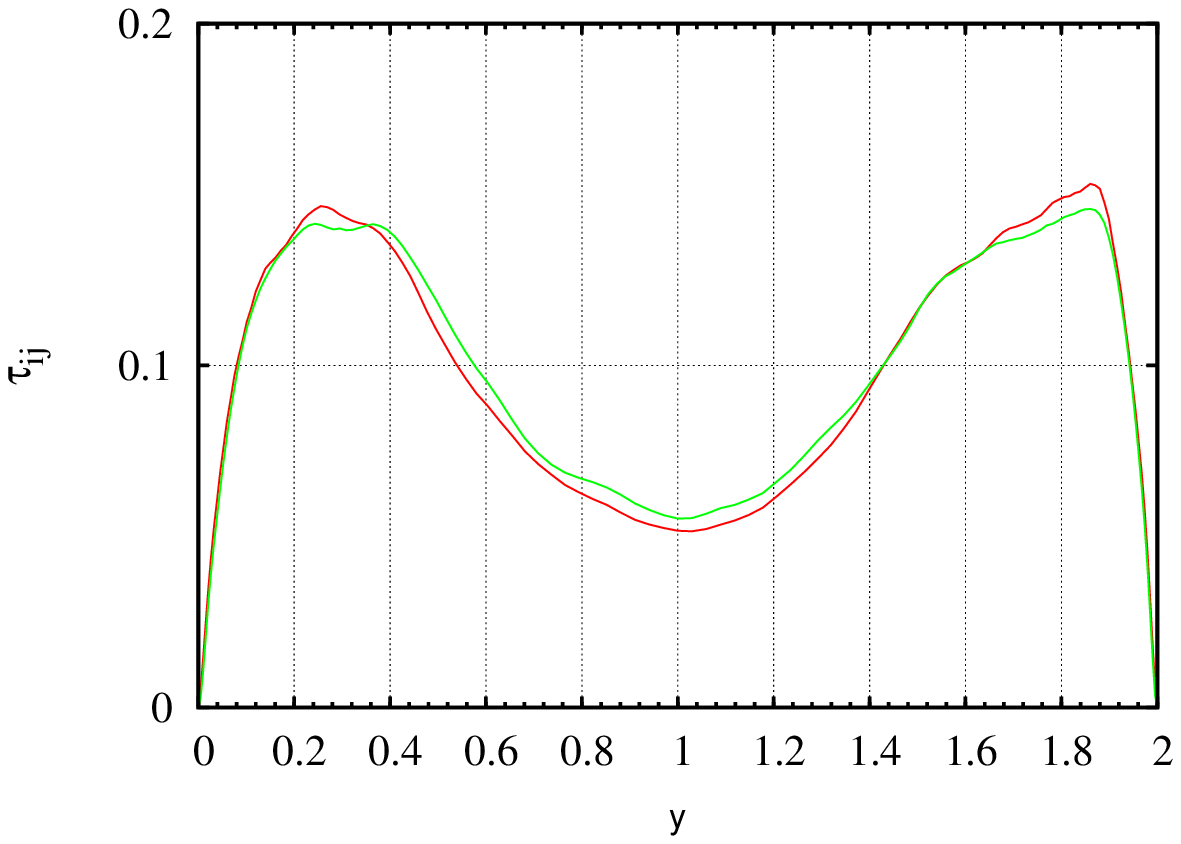}}
\end{minipage}
\hspace{5mm}
\begin{minipage}{0.45\hsize}
\begin{flushleft}
\hspace*{15mm}(d) $\tau_{12}$
\end{flushleft}
\vspace{-5mm}
   \centering
{\includegraphics[width=80mm]{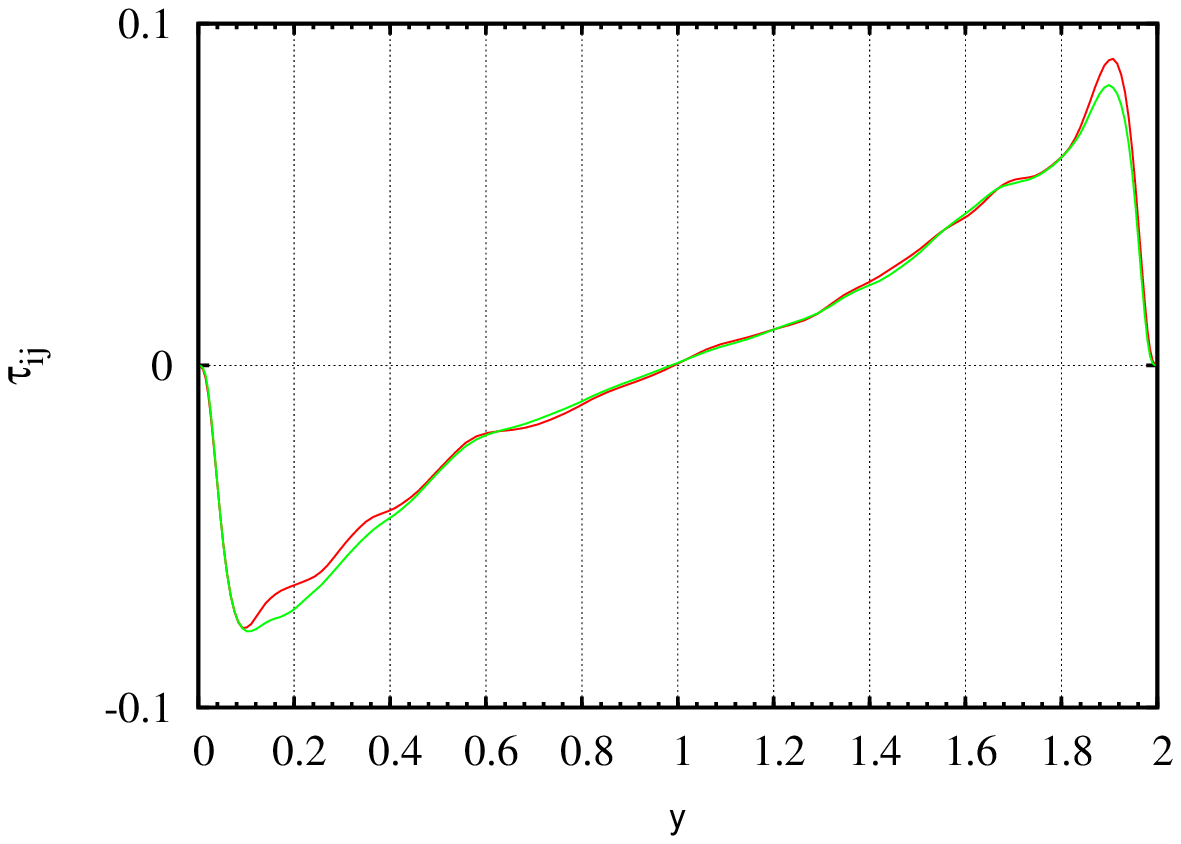}}
\end{minipage}
\vspace{5mm}

\begin{minipage}{0.45\hsize}
\begin{flushleft}
\hspace*{15mm}(e) $\tau_{23}$
\end{flushleft}
\vspace{-5mm}
   \centering
{\includegraphics[width=80mm]{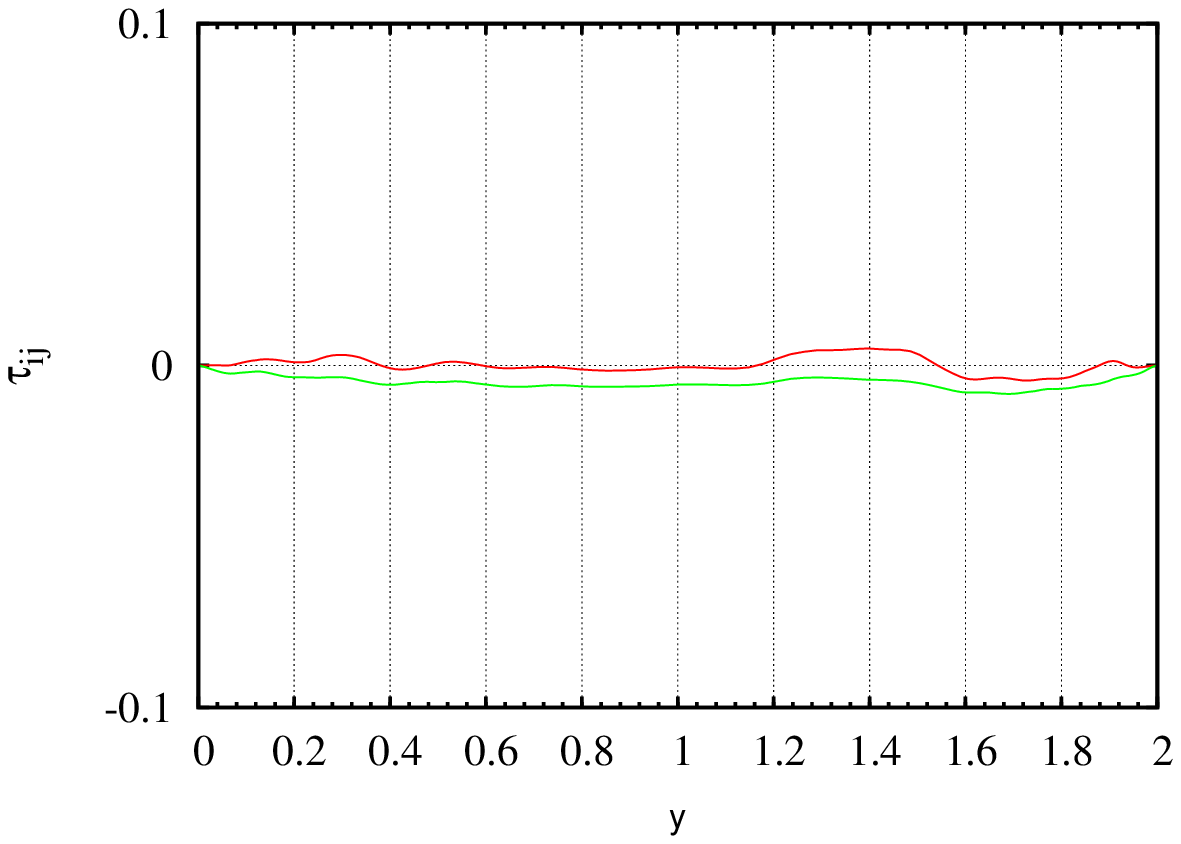}}
\end{minipage}
\hspace{5mm}
\begin{minipage}{0.45\hsize}
\begin{flushleft}
\hspace*{15mm}(f) $\tau_{31}$
\end{flushleft}
\vspace{-5mm}
   \centering
{\includegraphics[width=80mm]{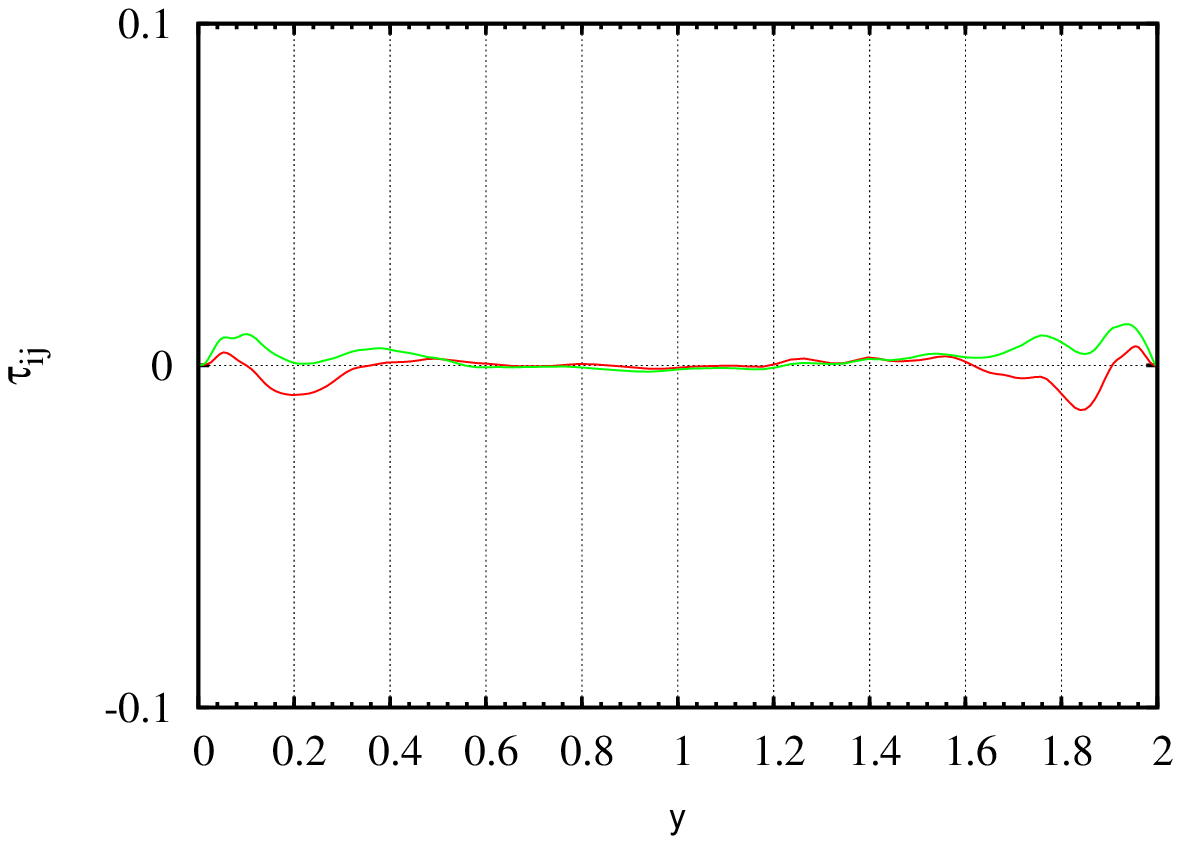}}
\end{minipage}
\caption{SGS stress averaged in streamwise and spanwise directions. 
Comparison between $\tau_{ij}^{(DNS)}$ and $\tau_{ij}^{(ANN)}$. 
$Re_\tau = 180, (\overline{\Delta x}^+, \overline{\Delta y}^+_{\rm{max}}, \overline{\Delta z}^+)=(35.3, 9.9, 17.7)$.}
\label{fig:Mean}
\end{figure}

\begin{figure}[h]
\begin{minipage}{0.45\hsize}
\begin{flushleft}
\hspace*{15mm}(a) $\tau_{11}$
\end{flushleft}
\vspace{-5mm}
   \centering
{\includegraphics[width=80mm]{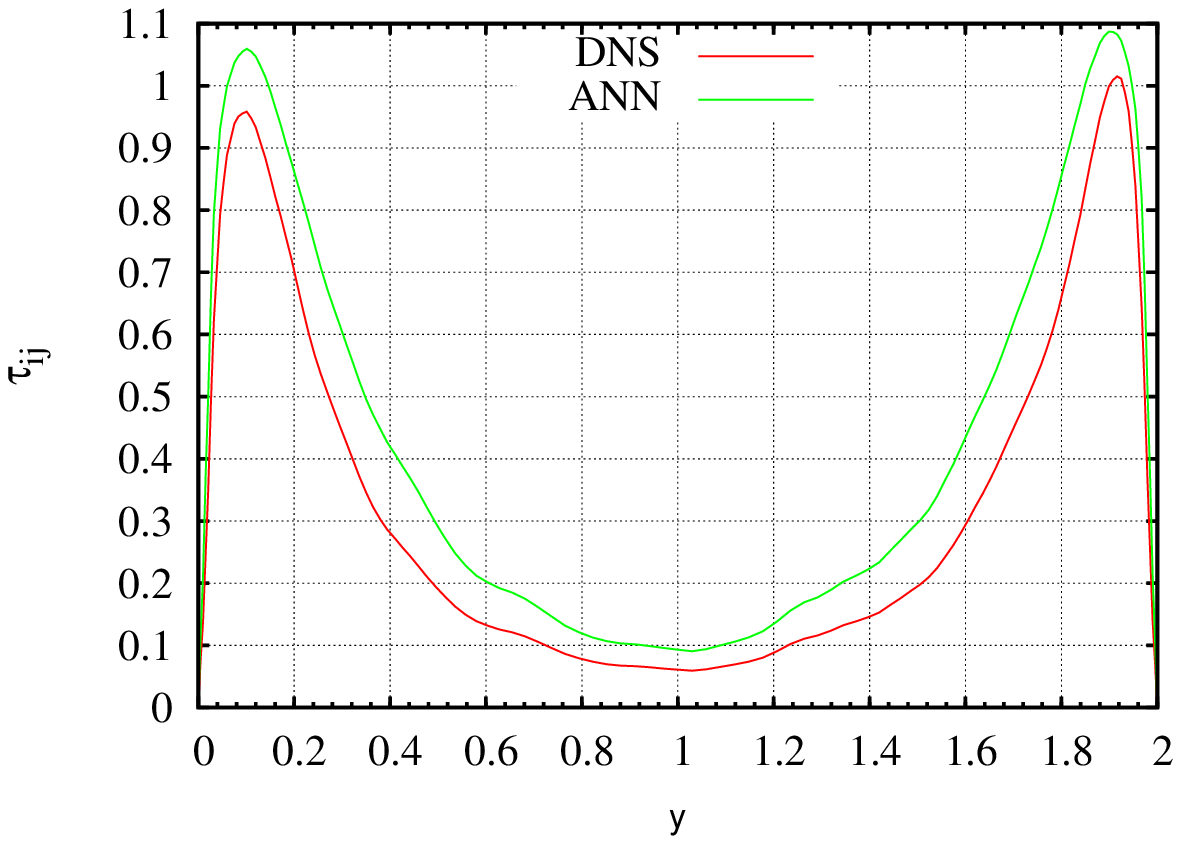}}
\end{minipage}
\hspace{5mm}
\begin{minipage}{0.45\hsize}
\begin{flushleft}
\hspace*{15mm}(b) $\tau_{22}$
\end{flushleft}
\vspace{-5mm}
   \centering
{\includegraphics[width=80mm]{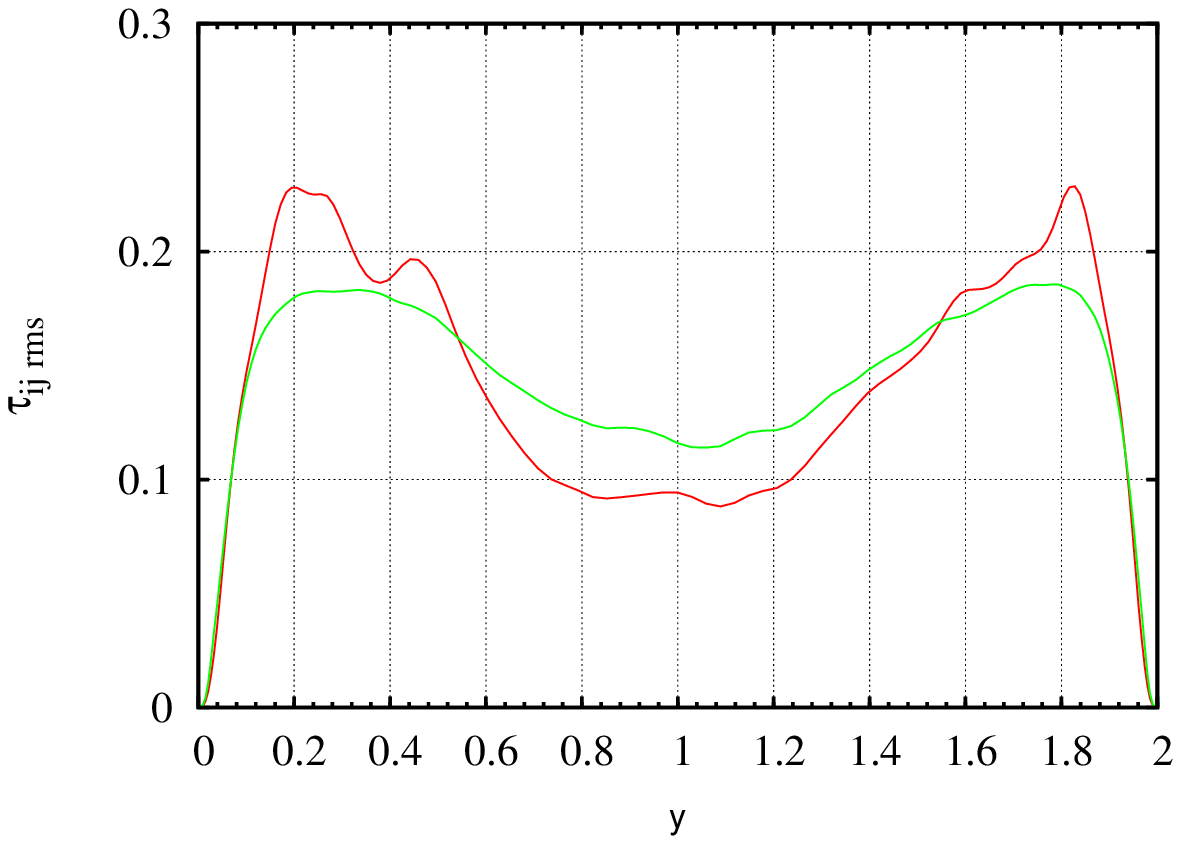}}
\end{minipage}
\vspace{5mm}

\begin{minipage}{0.45\hsize}
\begin{flushleft}
\hspace*{15mm}(c) $\tau_{33}$
\end{flushleft}
\vspace{-5mm}
   \centering
{\includegraphics[width=80mm]{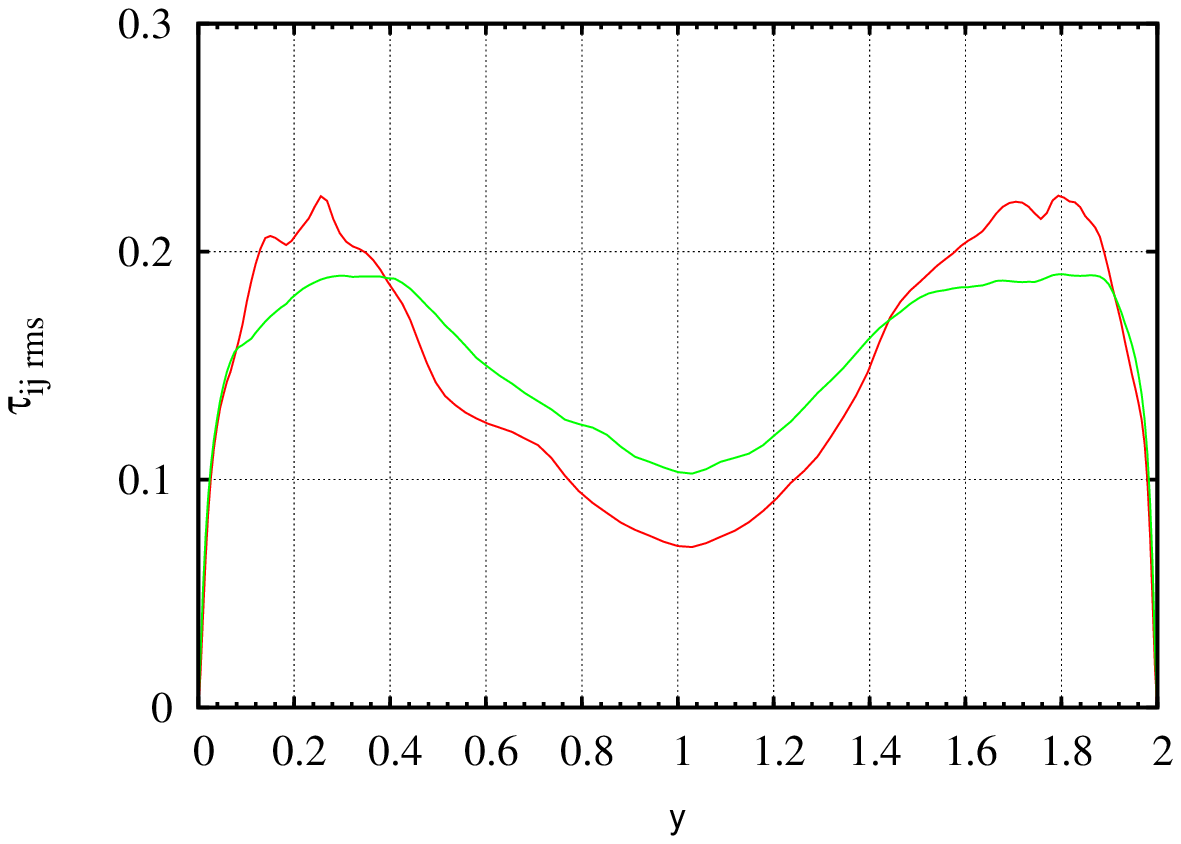}}
\end{minipage}
\hspace{5mm}
\begin{minipage}{0.45\hsize}
\begin{flushleft}
\hspace*{15mm}(d) $\tau_{12}$
\end{flushleft}
\vspace{-5mm}
   \centering
{\includegraphics[width=80mm]{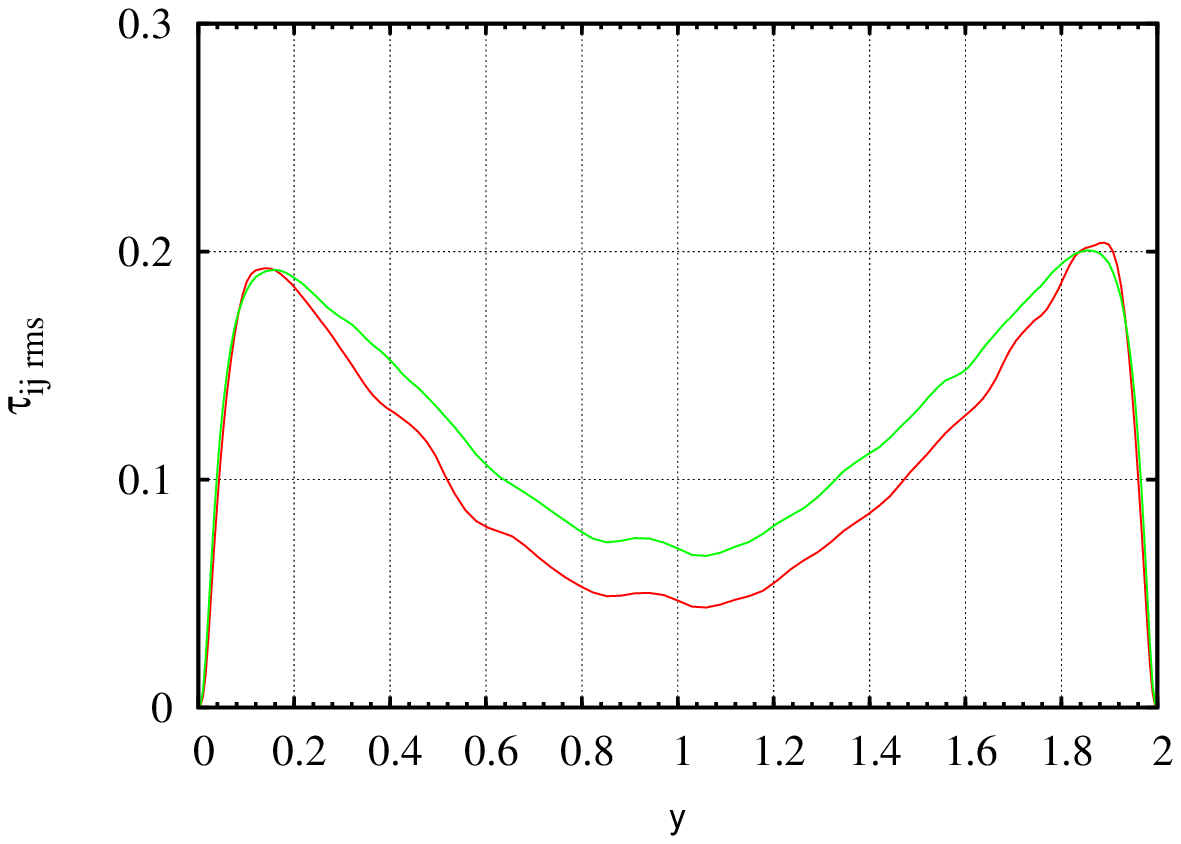}}
\end{minipage}
\vspace{5mm}

\begin{minipage}{0.45\hsize}
\begin{flushleft}
\hspace*{15mm}(e) $\tau_{23}$
\end{flushleft}
\vspace{-5mm}
   \centering
{\includegraphics[width=80mm]{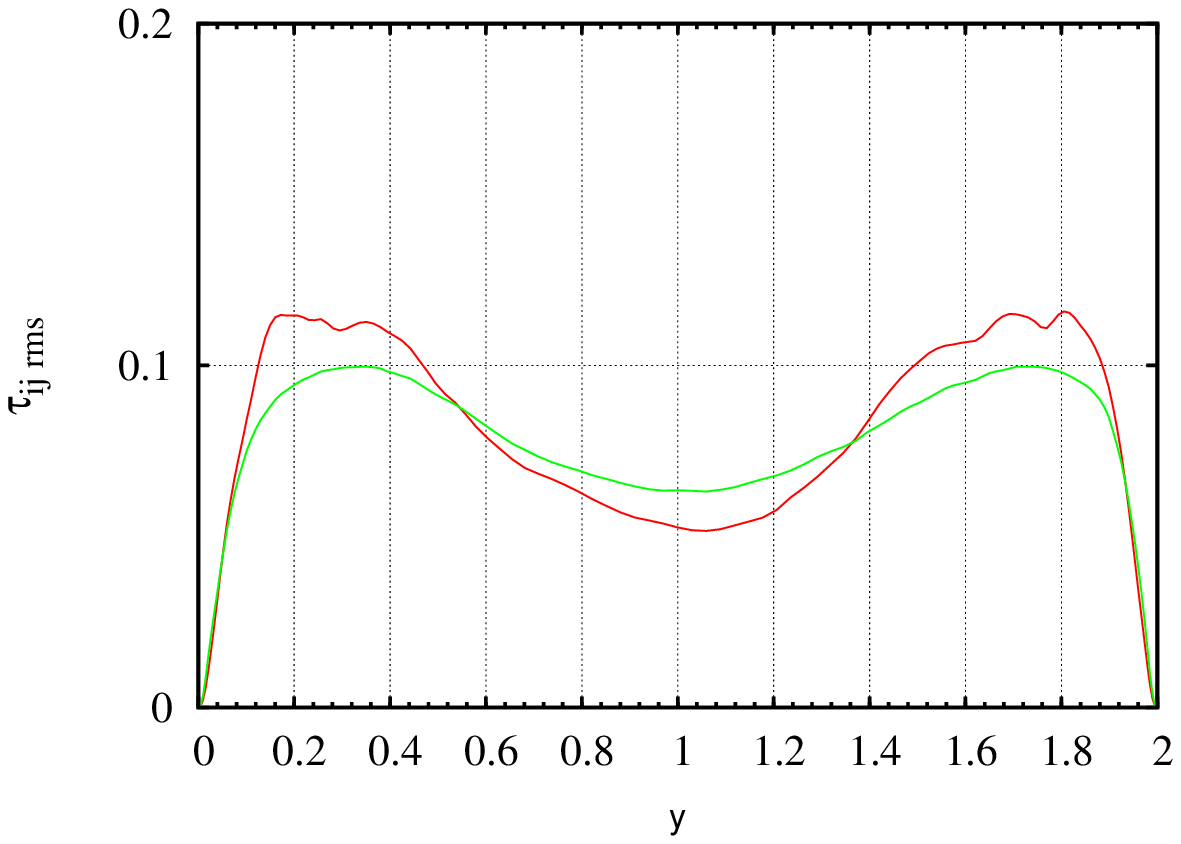}}
\end{minipage}
\hspace{5mm}
\begin{minipage}{0.45\hsize}
\begin{flushleft}
\hspace*{15mm}(f) $\tau_{31}$
\end{flushleft}
\vspace{-5mm}
   \centering
{\includegraphics[width=80mm]{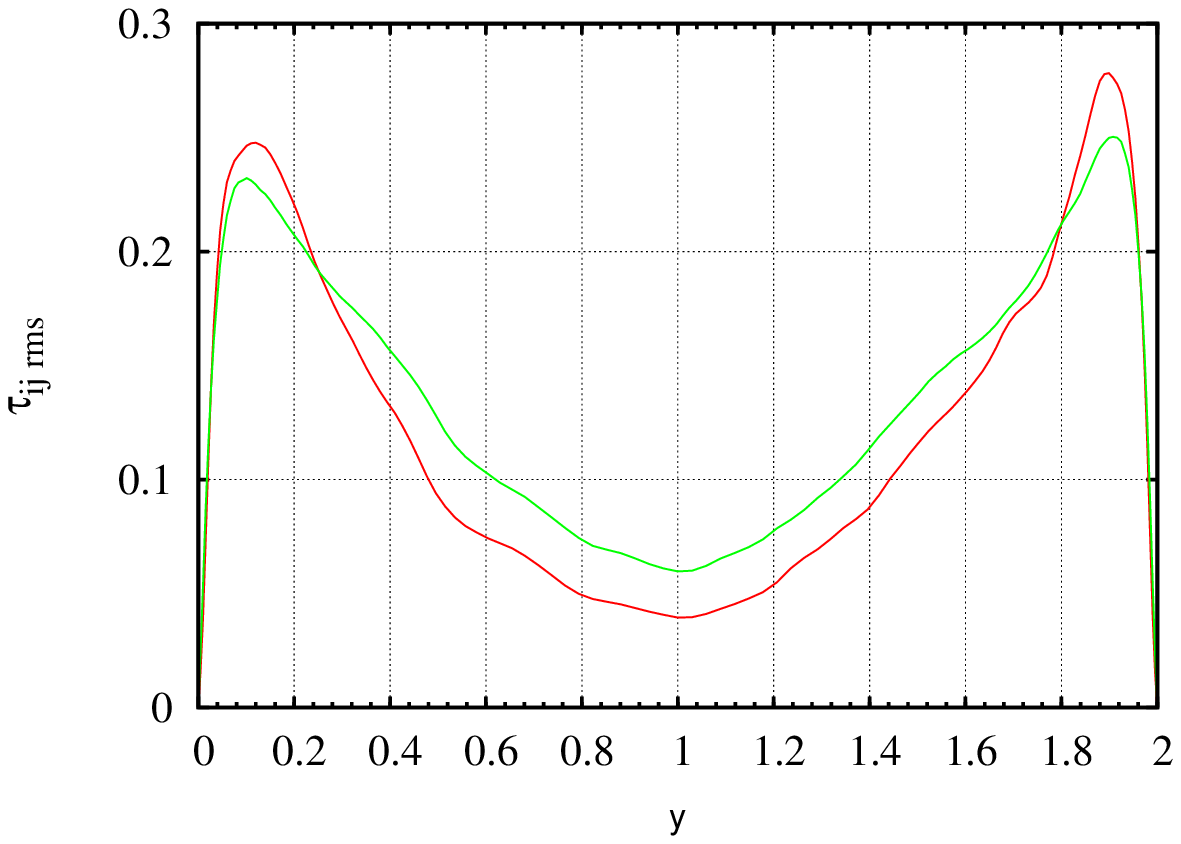}}
\end{minipage}
\caption{Rms amplitude of fluctuation of SGS stress in streamwise and spanwise directions. 
Comparison between $\tau_{ij}^{(DNS)}$ and $\tau_{ij}^{(ANN)}$. 
$Re_\tau = 180, (\overline{\Delta x}^+, \overline{\Delta y}^+_{\rm{max}}, \overline{\Delta z}^+)=(35.3, 9.9, 17.7)$.}
\label{fig:RMS}
\end{figure}

Next, we look into some details of the learning results for 
$Re_\tau = 180, (\overline{\Delta x}^+, \overline{\Delta y}^+_{\rm{max}}, \overline{\Delta z}^+)=(35.3, 9.9, 17.7)$. 
Figures \ref{fig:Dis} and \ref{fig:Dis2} compares the distributions of 
the SGS stress obtained by filtering DNS data, $\tau_{ij}^{(DNS)}$,   
and that predicted by trained ANN, $\tau_{ij}^{(ANN)}$. 
We choose the plane $y=0.1$ where the rms amplitudes of all components are nearly largest.  
The values of each component are normalized to be in $[0, 1]$. 
ANN is seen to reproduce fairly well the patterns in the distributions of DNS.   
Good agreement is also observed for the average (Fig. \ref{fig:Mean}) and rms amplitude 
(Fig. \ref{fig:RMS}) of each component. 
The learning is successful. 

\subsection{Basic features of learning}
\label{sec-res-basic}

In this subsection we show some features of learning by ANN 
which can be important in efficient applications. 
Table \ref{CC-table2} confirms that there is no significant 
dependence on the data set used in training. 
Five different data sets are tested. 
The difference between the correlation coefficients and the average is less than $0.043$. 

\begin{table}[h]
 	\begin{center}
    \vspace{5mm}
		\caption{Correlation coefficients between the SGS stress $\tau_{ij}^{(DNS)}$ calculated using DNS data 
and $\tau_{ij}^{(ANN)}$ predicted by trained ANN. 
Dependence on training data set. 
Correlation coefficients are averaged in the whole domain. }
		\def~{\hphantom{0}}
  		\begin{tabular}{ccccccc}
                  \hline\hline
			data set &$\tau_{11}$&$\tau_{22}$&$\tau_{33}$&$\tau_{12}$&$\tau_{23}$&$\tau_{31}$\\ \hline
			1st&0.796~&0.696~&0.750~&	0.760~&0.731~&0.826~\\
			2nd&0.802~&0.711~&0.693~&	0.753~&0.734~	&0.784~\\
			3rd&0.835~&0.715~&0.743~&	0.834~&0.750~	&0.832~\\
			4th&0.768~&0.709~&0.708~&	0.792~&0.732~	&0.826~\\
			5th&0.816~&0.732~	&0.747~&	0.815~&0.703~	&0.839~\\[-3pt] \hline
			Average&0.804~&0.713~&0.728~&	0.791~&0.730~&0.821~\\
                  \hline\hline
  		\end{tabular}
  		\label{CC-table2}
  	\end{center}
\end{table}

\begin{figure}[h]
	\centerline{\includegraphics[width=80mm]{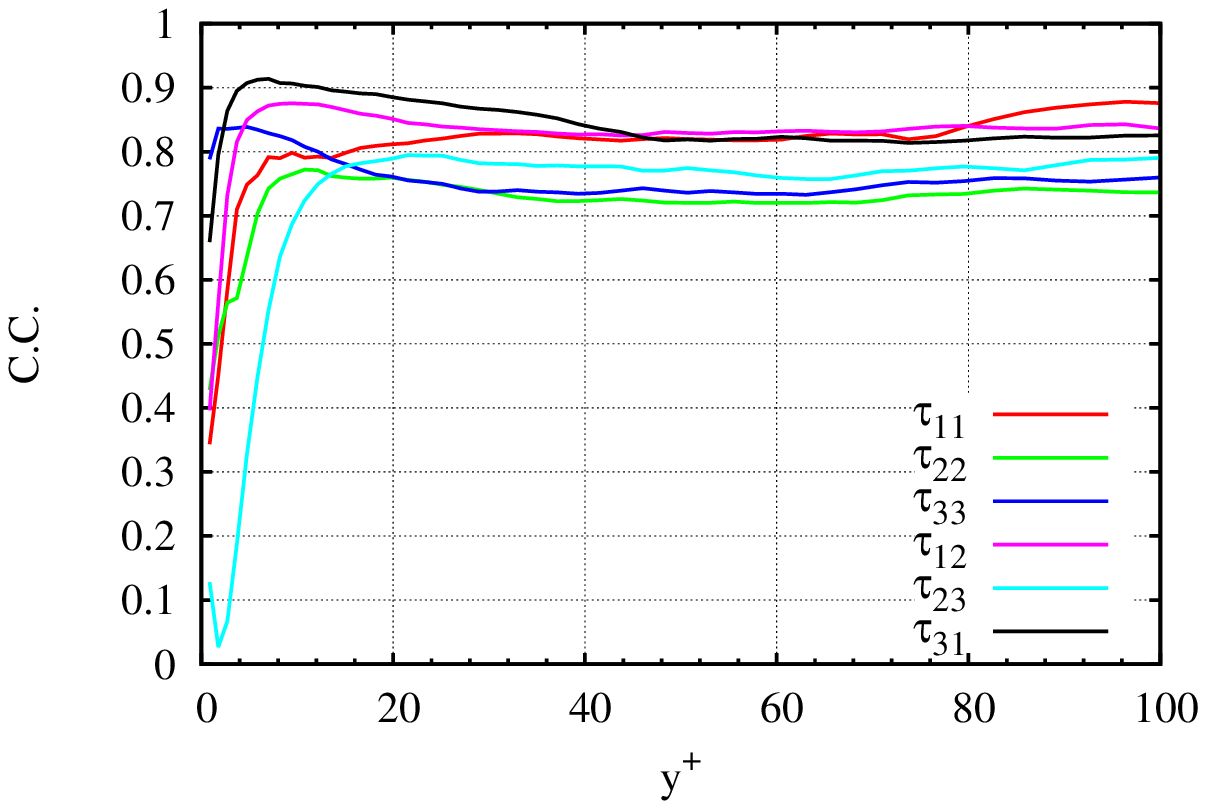}}   
  \vspace{3mm}
  	\caption{Correlation coefficients between $\tau_{ij}^{(DNS)}$  
and $\tau_{ij}^{(ANN)}$ near wall region. 
Correlation coefficients are averaged in streamwise and spanwise directions. }
	\label{fig:NearWall}
\end{figure}

Figure \ref{fig:NearWall} is a magnified view of Fig. \ref{fig:all-CC} near the wall. 
It shows that 
the correlation coefficients are small in the near wall region $y^+ < 10$, 
which consists of the viscous sublayer and a part of the buffer layer;  
it is reasonable since the flow field is not fully turbulent in this region. 
Outside this region, however, the correlation coefficients do not depend on the position $y$ significantly.  

\begin{table}[h]
	\begin{center}
    \vspace{5mm}
 		\caption{Correlation coefficients between $\tau_{ij}^{(DNS)}$  
and $\tau_{ij}^{(ANN)}$. 
Dependence on the number of neurons. 
Correlation coefficients are averaged in the whole domain. }
		\def~{\hphantom{0}}
		\begin{tabular}{ccccccc}
                  \hline\hline
			$n$ &$\tau_{11}$&$\tau_{22}$&$\tau_{33}$&$\tau_{12}$&$\tau_{23}$&$\tau_{31}$\\ \hline
    	2~&0.677~&	0.487~&	0.486~&	0.366~&	0.398~&	0.547~\\
	    5~&0.752~&	0.671~&	0.674~&	0.593~&	0.604~&	0.710~\\
	    10~&0.742~&	0.672~&	0.717~&	0.779~&	0.711~&	0.801~\\
	    25~&0.782~&	0.684~&	0.739~&	0.807~&	0.736~&	0.807~\\
	    50~&0.806~&	0.709~&	0.739~&	0.796~&	0.739~&	0.812~\\
	    100~&0.827~&	0.711~&	0.739~&	0.788~&	0.736~&	0.816~\\
                  \hline\hline
  		\end{tabular}
  		\label{tab:Neuron}
  	\end{center}
\end{table}

\begin{figure}[h]
  \begin{tabular}{cc}
    \begin{minipage}{0.5\hsize}
      \begin{center}
{\includegraphics[width=80mm]{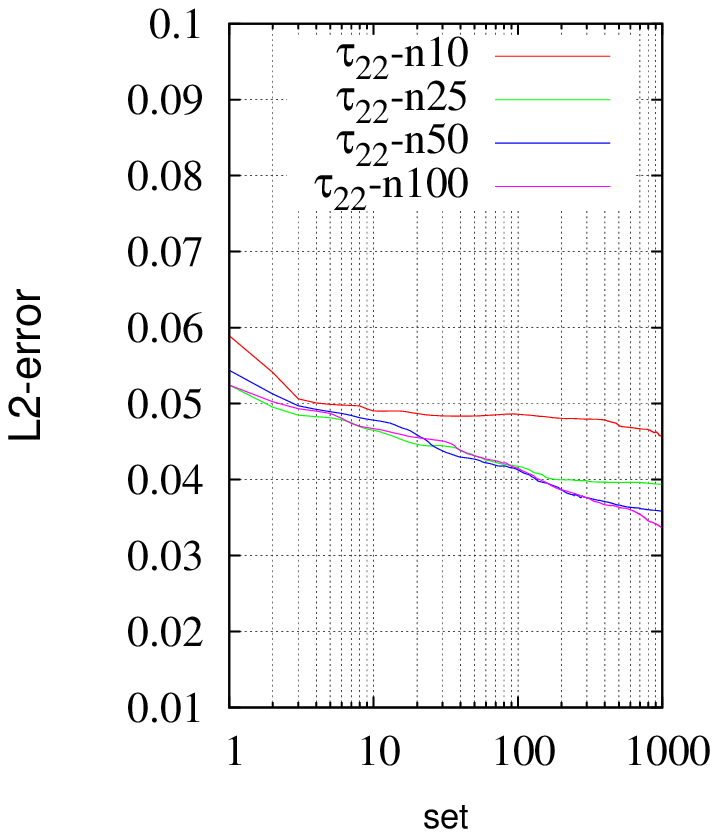}}
      \end{center}
    \end{minipage}
    \begin{minipage}{0.5\hsize} 
      \begin{center}
{\includegraphics[width=80mm]{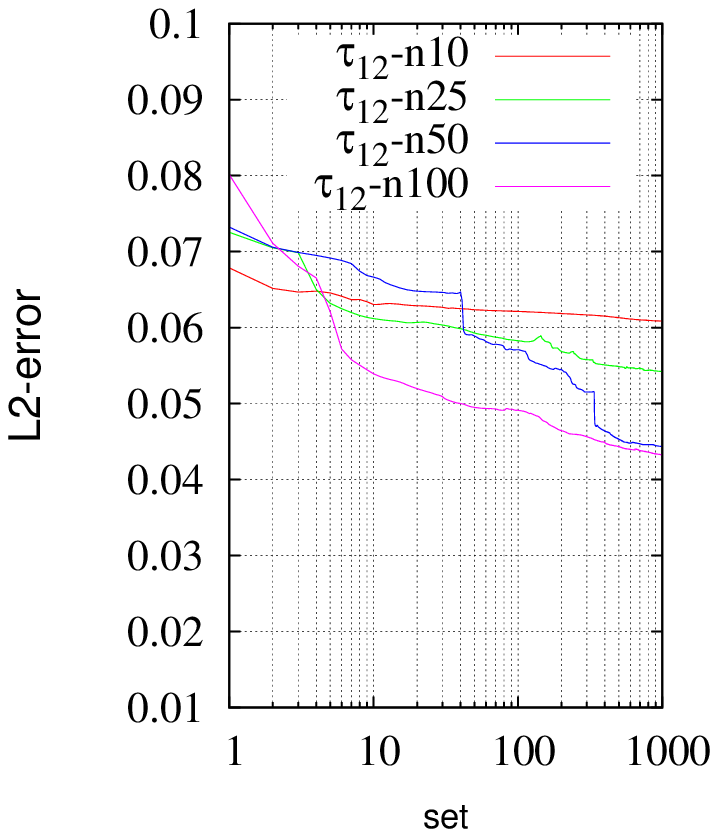}}
      \end{center}
    \end{minipage}
  \end{tabular}
  \vspace{5mm}
  \caption{$L^2$-norm of error between the output and the training target 
as a function of number of iterations. 
(Left) $\tau_{12}$, (right) $\tau_{12}$. }
  \label{fig:N-error}
\end{figure}

Table \ref{tab:Neuron} shows how many neurons are needed for successful learning. 
In this table $n$ is the number of neurons in the hidden layer.  
In general, approximation by ANN becomes more accurate by increasing $n$  
with the expense of computational time.  
It is seen that the correlation coefficients increase with $n$, 
but the increase is slow for $n \ge 10$. 
All correlation coefficients exceed $0.7$ when $n \ge 50$. 
Figure \ref{fig:N-error} shows how the difference between DNS and ANN 
decays with the number of iterations in training. 
As the number of neurons increases the rate of decay increases   
and the final error after $1000$ iterations of learning decreases. 
On the other hand, 
the time required for ANN calculation increases with $n$. 
Thus the choice of $n$ should be made taking account of 
the accuracy of approximation as well as the calculation time. 

\begin{figure}[h]
	\centerline{\includegraphics[width=80mm]{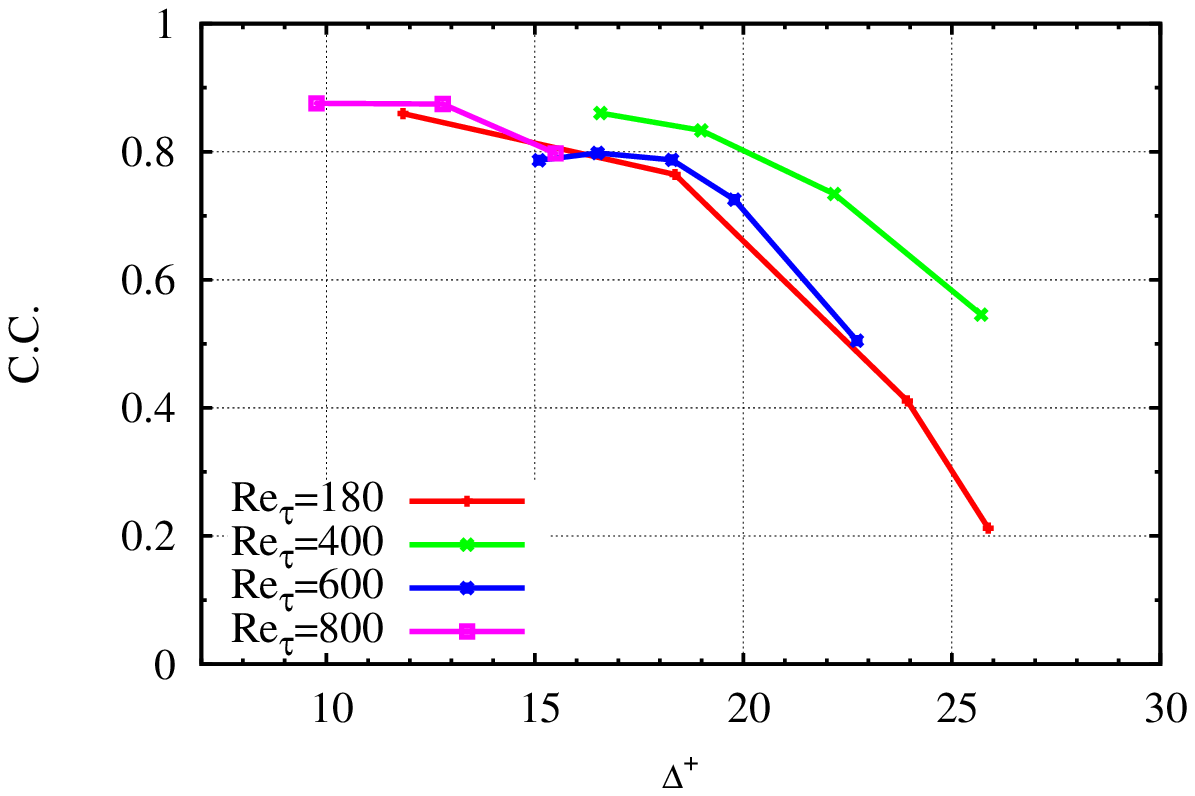}}
  \vspace{5mm}
 	\caption{Correlation coefficients between $\tau_{ij}^{(DNS)}$  
and $\tau_{ij}^{(ANN)}$. 
Dependence on the filter size. 
Correlation coefficients are averaged in the whole domain. }
	\label{fig:Reynolds}
\end{figure}

Figure \ref{fig:Reynolds} shows the overall correlation coefficients 
as a function of the averaged filter size 
$\overline{\Delta}^+=\left(\overline{\Delta x}^+\overline{\Delta y}^+_{\rm{max}}\overline{\Delta z}^+\right)^{1/3}$ 
for various Reynolds numbers. 
It is seen that learning is successful for $\Delta^+ \lesssim 20$, 
which corresponds to $3 \sim 4$ times grid spacing. 
The correlation coefficients decrease quickly as the filter size becomes large.  
Thus there is a limitation for the filter size.  
There is little dependence on the Reynolds number, 
although the correlation coefficient is larger for $Re_\tau=400$ than 
the other three cases.  

\subsection{Applicability to higher Reynolds numbers}

\begin{figure}[h]
\begin{center}
{\includegraphics[width=80mm]{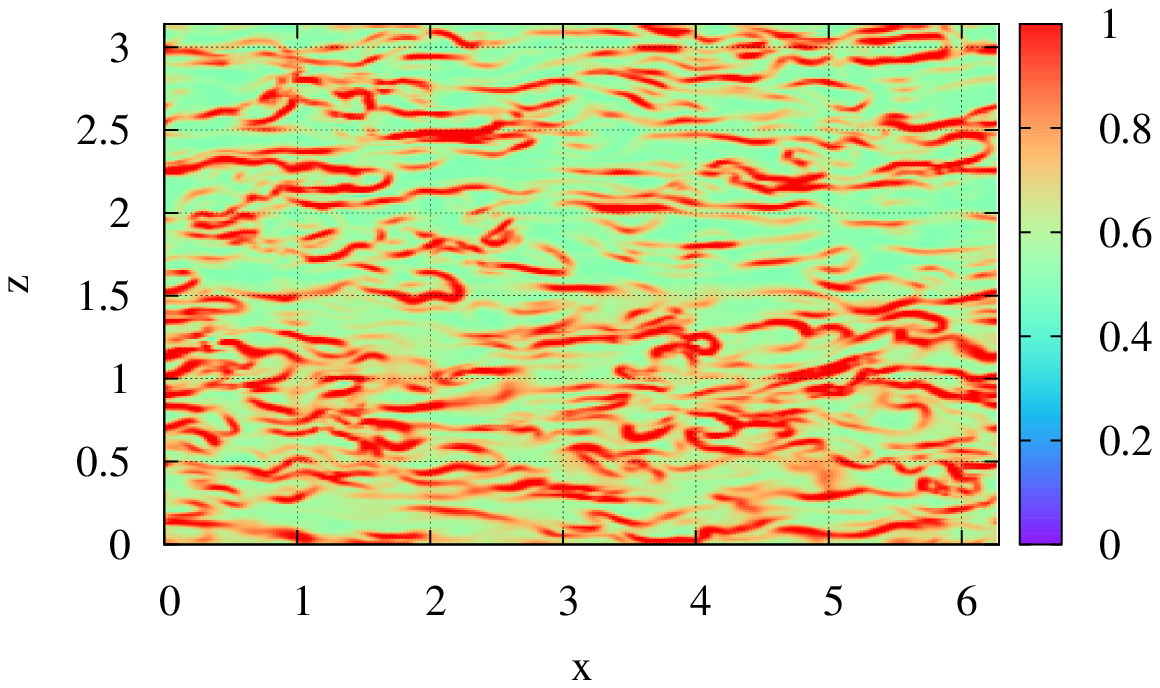}}
{\includegraphics[width=80mm]{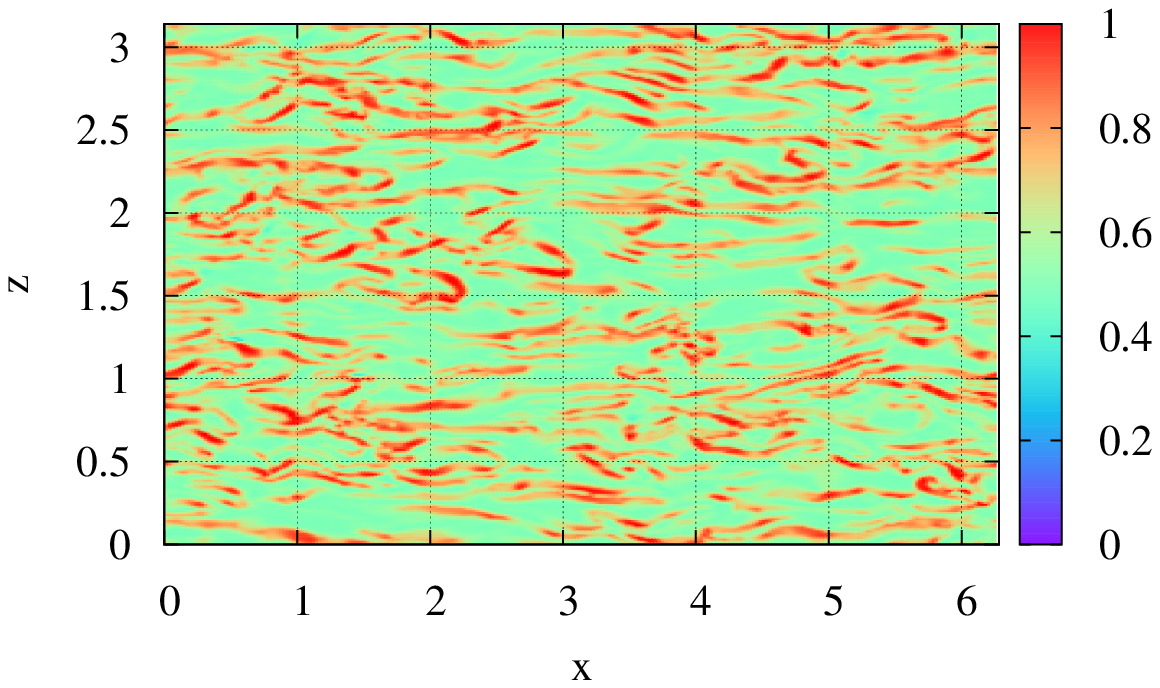}}
  \caption{Comparison between $\tau_{11}^{(DNS)}$ obtained by filtering DNS data 
of $Re_\tau=400$ with filter size $(\overline{\Delta x}^+, \overline{\Delta y}^+_{\rm{max}}, \overline{\Delta z}^+)=(34.3, 17.9, 17.5)$ 
and $\tau_{11}^{(ANN)}$ predicted by ANN trained at $Re_\tau=180$ with filter size 
$(\overline{\Delta x}^+, \overline{\Delta y}^+_{\rm{max}}, \overline{\Delta z}^+)=(35.3, 9.9, 17.7)$.  
$y=0.1. $}
	\label{fig:extrapolation}
\end{center}
\end{figure}

LES is usually used for high Reynolds number flows 
which are not accessible by DNS.  
Thus when we think of applying ANN to LES of high Reynolds number flows,  
no training data are available from DNS. 
In this regard, 
it is of importance to check 
whether ANN trained at low Reynolds numbers is useful at high Reynolds numbers. 
This is checked applying ANN trained at $Re_\tau=180$ to 
predicting the SGS stress tensor at $Re_\tau=400$. 
Figure \ref{fig:extrapolation} 
compares the distribution of $\tau_{11}^{(DNS)}$ obtained by filtering DNS data 
of $Re_\tau=400$ with filter size $(\overline{\Delta x}^+, \overline{\Delta y}^+_{\rm{max}}, \overline{\Delta z}^+)=(34.3, 17.9, 17.5)$ 
and $\tau_{11}^{(ANN)}$ predicted by ANN trained at $Re_\tau=180$ with filter size 
$(\overline{\Delta x}^+, \overline{\Delta y}^+_{\rm{max}}, \overline{\Delta z}^+)=(35.3, 9.9, 17.7)$.  
Although there is some difference in the magnitude,  
the spatial patterns in DNS data are reproduced by ANN. 
The correlation coefficients between DNS and ANN are 
larger than $0.7$ (Table \ref{tab:extrapolation}).  
These results support that 
it is possible to use ANN trained at low Reynolds numbers 
for LES at high Reynolds numbers. 

\begin{table}[h]
	\begin{center}
    \vspace{5mm}
		\caption{Correlation coefficients between $\tau_{ij}^{(DNS)}$ obtained by filtering DNS data 
of $Re_\tau=400$ with filter size $(\overline{\Delta x}^+, \overline{\Delta y}^+_{\rm{max}}, \overline{\Delta z}^+)=(34.3, 17.9, 17.5)$ 
and $\tau_{ij}^{(ANN)}$ predicted by ANN trained at $Re_\tau=180$ with filter size 
$(\overline{\Delta x}^+, \overline{\Delta y}^+_{\rm{max}}, \overline{\Delta z}^+)=(35.3, 9.9, 17.7)$. 
Correlation coefficients are averaged in the whole domain. }

		\def~{\hphantom{0}}
		\begin{tabular}{cccccc}
                  \hline\hline
			$\tau_{11}$&$\tau_{22}$&$\tau_{33}$&$\tau_{12}$&$\tau_{23}$&$\tau_{31}$\\ \hline
			0.707~&	0.744~&	0.734~&	0.743~&	0.768~&	0.769~\\
                  \hline\hline
  		\end{tabular}
  		\label{tab:extrapolation}
  	\end{center}
\end{table}

\subsection{What ANN has learned?}

Finally, we investigate what kind of model ANN has established. 
First, we identify and eliminate members of the input variables 
which are not required for ANN approximation 
for each component of the SGS stress tensor. 
This is done by removing components of $\pmb{\nabla}\overline{\pmb{u}}$ one by one 
and checking whether the correlation coefficients become much smaller than before. 
Table \ref{tab:CC-table-mat} shows the results;  
the members of the input variables 
required for achieving high correlation coefficients 
are marked by circles. 
The values of correlation coefficients with the reduced numbers of input variables 
are shown in the first row of Table \ref{tab:CC-table3}. 
They are larger than those obtained with the full members of $\{\mathbd{\nabla} \overline{\mathbd{u}}, y\}$ 
(Table \ref{CC-input-table1}). 
Thus the results of learning are improved by eliminating irrelevant components. 

\begin{table}[h]
 	\begin{center}
    \vspace{5mm}
		\caption{Input variables required for accurate prediction by ANN. }
		\def~{\hphantom{0}}
  		\begin{tabular}{cccccccccc}
                  \hline\hline
& $\pd{u}{x}$ & $\pd{u}{y}$ & $\pd{u}{z}$ & $\pd{v}{x}$ & $\pd{v}{y}$ & $\pd{v}{z}$ & $\pd{w}{x}$ & $\pd{w}{y}$ & $\pd{w}{z}$ \\ 
                  \hline
$\tau_{11}$ & $\circ$ & & $\circ$ & & & & & & \\
$\tau_{22}$ & & & & $\circ$ & & $\circ$ & & & \\
$\tau_{33}$ & & & & & & & $\circ$ & & $\circ$ \\
$\tau_{12}$ & $\circ$ & & $\circ$ & & & & & & \\
$\tau_{23}$ & & & & $\circ$ & & $\circ$ & $\circ$ & & $\circ$ \\
$\tau_{31}$ & $\circ$ & & $\circ$ & & & & $\circ$ & & $\circ$ \\
                  \hline\hline
  		\end{tabular}
  		\label{tab:CC-table-mat}
  	\end{center}
\end{table}

Next, we infer the model ANN has produced. 
The above results imply that 
\begin{eqnarray}
\tau_{ij} = f \left(\pd{\overline{u}_i}{x}, \pd{\overline{u}_i}{z}, \pd{\overline{u}_j}{x}, \pd{\overline{u}_j}{z}, y\right),   
\end{eqnarray}
which reminds us of the gradient model 
\begin{eqnarray}
\tau_{ij} = \frac{\overline{\Delta}^2}{12} \pd{\overline{u}_i}{x_k}\pd{\overline{u}_j}{x_k}. 
\end{eqnarray}
We prefer the following form 
\begin{eqnarray}
\tau_{ij} = \sum_{k=1}^3 \frac{\overline{\Delta_k}^2}{12} \pd{\overline{u}_i}{x_k}\pd{\overline{u}_j}{x_k},  
\label{eq:grad}
\end{eqnarray}
since the grid spacings in $y$ direction are much smaller than the other two directions.  

\begin{figure}[h]
\begin{center}
{\includegraphics[width=80mm]{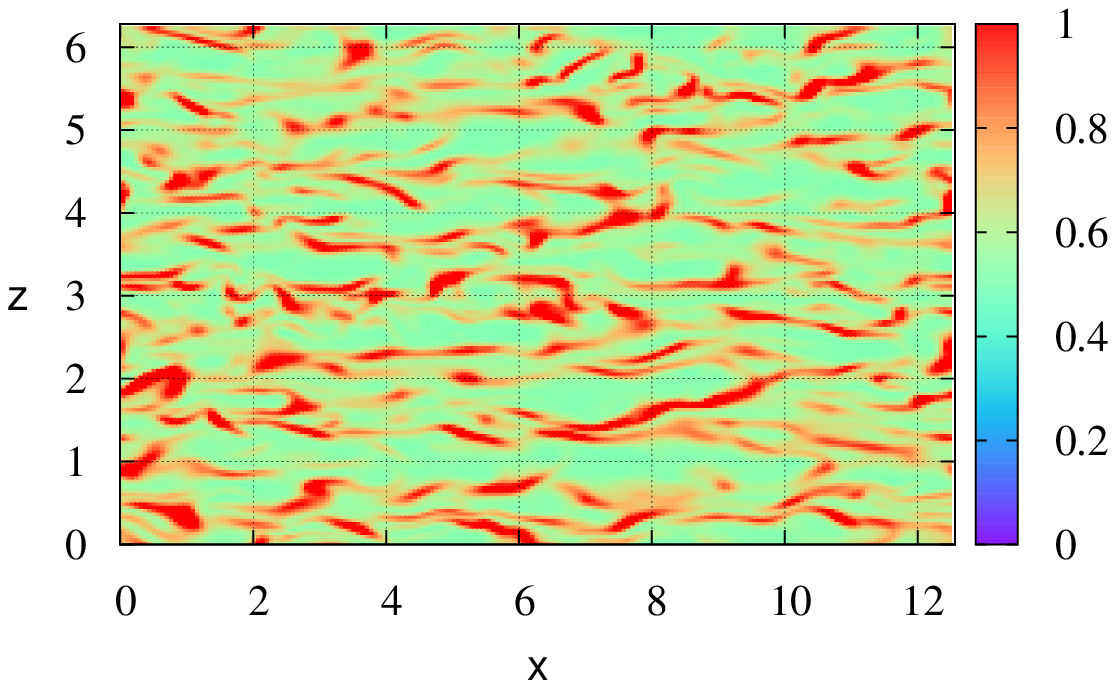}}
  \caption{Distribution of $\tau_{11}$ obtained by gradient model. 
$Re_\tau=180$, $(\overline{\Delta x}^+, \overline{\Delta y}^+_{\rm{max}}, \overline{\Delta z}^+)=(35.3, 9.9, 17.7)$. }
	\label{fig:grad}
\end{center}
\end{figure}

Figure \ref{fig:grad} shows the distribution of $\tau_{11}$ obtained by the gradient model. 
It is similar to not only $\tau_{11}^{(DNS)}$ but also $\tau_{11}^{(ANN)}$ (Fig. \ref{fig:Dis}). 
In Table \ref{tab:CC-table3} the correlation 
between the gradient model (\ref{eq:grad}) and DNS 
is high and comparable to the correlation between ANN and DNS. 
The correlation coefficients between the gradient model and ANN 
are a bit smaller than those in the other two rows, 
but still large. 
Thus, ANN has established a model which is similar to the gradient model. 


\begin{table}[h]
 	\begin{center}
    \vspace{5mm}
		\caption{Correlation coefficients between (top) $\tau_{ij}^{(DNS)}$ and 
$\tau_{ij}^{(ANN)}$ with reduced number of input variables, 
(middle) the gradient model and $\tau_{ij}^{(DNS)}$,  
and (bottom) the gradient model and $\tau_{ij}^{(ANN)}$.  
Correlation coefficients are averaged in the whole domain. 
$Re_\tau=180$, $(\overline{\Delta x}^+, \overline{\Delta y}^+_{\rm{max}}, \overline{\Delta z}^+)=(35.3, 9.9, 17.7)$. 
}
		\def~{\hphantom{0}}
  		\begin{tabular}{ccccccc}
                  \hline\hline
			Case &$\tau_{11}$&$\tau_{22}$&$\tau_{33}$&$\tau_{12}$&$\tau_{23}$&$\tau_{31}$\\ \hline
		 	reduced input variables &0.863~&0.769~&0.793~&0.848~&0.793~&0.856~\\
		 	gradient model vs. DNS &0.847~&0.729~&0.737~&0.841~&0.703~&0.853~\\
		 	gradient model vs. ANN &0.651~&0.675~&0.634~&0.758~&0.664~&0.640~\\
                  \hline\hline
  		\end{tabular}
  		\label{tab:CC-table3}
  	\end{center}
\end{table}

\section{Concluding Remarks}

We have shown that 
ANN can establish a functional relation between 
the GS flow field and the SGS stress tensor in LES. 
Using DNS data of a turbulent channel flow as training data, 
ANN was trained by back propagation. 
Then the ability of the trained ANN was tested 
using DNS data which were not used in training. 
Learning was most successful 
when $\{\mathbd{\nabla} \overline{\mathbd{u}}, y\}$ was used as a set of the input variables; 
the correlation coefficients between the SGS stress tensor obtained by 
filtering DNS data and that predicted by ANN exceeded $0.7$. 
The spatial distribution of the SGS stress tensor predicted by ANN 
was in good agreement with that obtained by filtering DNS data. 
It is most likely that ANN has established a model close to the gradient model. 

It should be pointed out that 
learning was successful only when the filter size was small: $\overline{\Delta}^+ \lesssim 20$. 
For this small filter size the SGS Reynolds stress term is so small that 
the SGS stress is close to 
the sum of the Leonard term and the cross-stress term, 
which can be approximated by the gradient model. 
Therefore, the present results are quite reasonable. 
However, it should be stressed that 
ANN has succeeded in establishing a functional relation between the GS flow field 
and the SGS stress without any assumption on the form of function. 
We also remark that 
the values of correlation coefficients between DNS and ANN obtained in the present study 
are comparable to or even larger than those for similarity models \cite{H-1989, LMK-1994}  
but smaller than those for a dynamic two-parameter mixed model \cite{H-1997},  
which implies that ANN is a promising tool for searching for new turbulence models 
and should be improved. 

In the present study we have not taken care of symmetry of the SGS stress tensor; 
each component of the tensor is trained separately.   
If the fluctuations are statistically isotropic, however, 
the turbulence model should be also isotropic;   
training is required for 
only one diagonal component and one off-diagonal component of the SGS stress tensor, 
while the two are related through the incompressibility condition. 
In the present study, however, the grid spacings and the filter size differ 
depending on the direction; 
this is one of the reasons why each component of the SGS stress tensor 
is trained independently. 
It should be noted that it is important to take account of symmetry 
when we formulate a trained ANN into a turbulence model. 

Although we have obtained some successful results with ANN, 
several problems should be solved in order to 
find a new turbulence model better than the existing ones, 
which is our final goal.  
We would like to find a model which works well even when 
the filter size is much larger than the grid size. 
However, 
it may be impossible to have high correlation of the SGS stress tensor between ANN and DNS 
for large filter size 
since SGS fluctuations involve a wide range of length scales and are inevitably large. 
In this regard, one way to proceed is to allow a certain level of errors, 
which can be incorporated in other methods of machine learning like support vector machine/regression. 
Another way is to replace the training target or output variable; 
in the present study we chose the SGS stress tensor as the output variable, 
but it can be other quantities like the rate of production of residual energy or SGS dissipation, 
which is important in the energy transfer between GS and SGS scales.  

ANN itself can be also improved. 
There are a number of possibilities in the choice of input variables. 
In the present study the input variables were chosen taking account of the Smagorinsky model, 
but it can be replaced by the similarity model and we can include doubly-filtered variables 
in the input variables. 
We may include all data if the ability of ANN allows it. 
In this regard, tuning of ANN would be also important;  
the number of layers may be increased as we see the remarkable success of deep learning.



%
%

%

\begin{acknowledgments}
Numerical calculations were performed on the Altix UV1000 and UV2000 at the 
Advanced Fluid Information Research Center, 
Institute of Fluid Science, Tohoku University.
\end{acknowledgments}


\end{document}